\RequirePackage{fix-cm}
\documentclass{svjour3}                     
\smartqed  
\usepackage{amsmath}
\usepackage{amssymb}
\usepackage{natbib}
\bibpunct{(}{)}{;}{a}{}{,~}
\usepackage{graphicx}
\usepackage[labelfont=bf,labelsep=space]{caption}
\usepackage{dsfont}
\usepackage{booktabs}
\usepackage[utf8]{inputenc}
\usepackage{wrapfig,framed,caption}
\DeclareGraphicsExtensions{.eps}
\usepackage{xcolor}

\journalname{Bulletin of Mathematical Biology}

\begin{document}

\title{A mathematical model of the effects of aging on naive T-cell population
  and diversity}



\author{Stephanie Lewkiewicz \and Yao-li Chuang \and Tom Chou}


\institute{Stephanie Lewkiewicz \at
              \email{slewkiewicz@math.ucla.edu} \\
              Department of Mathematics, UCLA, Los Angeles, CA 90095-1555, USA
           \and
           Yao-li Chuang \at
    \email{ylch07@gmail.com} \\
           Department of Mathematics, CalState-Northridge, Northridge, CA 91330-8313, USA\\
           Department of Biomathematics, UCLA, Los Angeles, CA 90095-1766, USA
           \and
           Tom Chou \at
 \email{tomchou@ucla.edu} \\
           Department of Biomathematics, UCLA, Los Angeles, CA 90095-1766, USA\\
           Department of Mathematics, UCLA, Los Angeles, CA 90095-1555, USA}

\date{Received: date / Accepted: date}

\maketitle

\begin{abstract}
The human adaptive immune response is known to weaken in advanced age,
resulting in increased severity of pathogen-born illness, poor vaccine
efficacy, and a higher prevalence of cancer in the elderly.
Age-related erosion of the T-cell compartment has been implicated as a
likely cause, but the underlying mechanisms driving this
immunosenescence have not been quantitatively modeled and
systematically analyzed.  T-cell receptor diversity, or the extent of
pathogen-derived antigen responsiveness of the T-cell pool, is known
to diminish with age, but inherent experimental difficulties preclude
accurate analysis on the full organismal level.  In this paper, we
formulate a mechanistic mathematical model of T-cell population
dynamics on the immunoclonal subpopulation level, which provides
quantitative estimates of diversity.  We define different estimates
for diversity that depend on the individual number of cells in a
specific immunoclone.  We show that diversity decreases with age
primarily due to diminished thymic output of new T-cells and the
resulting overall loss of small immunoclones.
\end{abstract}

\newpage

\section{Introduction}
\label{intro}
Immunosenescence underlies poor health outcomes in the aging
population, including diminished vaccine
efficacy~(\citealt{POLAND2010,MCELHANEY2008,FLEMING2005}), increased
susceptibility to disease (including irregular presentation,
intensified symptoms, longer recovery times, and increased
mortality)~(\citealt{THOMAS2012}), as well as a heightened risk of
cancer~(\citealt{GINALDI2001}).  This degradative aging process of the
human immune system originates from extensive fundamental changes to
the size and functionality of immune cell pools, and the structure of
lymphatic tissues in which they develop and
operate~(\citealt{SALAM2013}).

Among the many changes associated with immunosenescence
(\citealt{GLOBERSON2000}), the T-cell compartment is arguably the most
damaged~(\citealt{WICK2000,GRUVER2007}).  The T-cell pool is comprised
of subpopulations of antigen-inexperienced naive cells, and
antigen-experienced memory cells, the latter of which retain
immunological record of previous infections.  The human immune
compartment maintains $\sim 10^{12}$ T-cells in total, of which $\sim
10^{11}$ are naive~(\citealt{JENKINS2009,TREPEL1974}).  During aging,
the population of naive T-cells declines in overall size, while the
population of memory T-cells undergoes extensive proliferation,
thereby reversing the balance of naive and memory T-cells that had
persisted at younger ages~(\citealt{GLOBERSON2000,FAGNONI2000}).  The
expansion of memory T-cells further enhances immunological memory of
previously-encountered antigens, reinforcing existent immune
protection.  The remaining naive pool experiences loss of T-cell
receptor (TCR) ``structural
diversity"~(\citealt{GORONZY2007,GORONZY2015})--the number of distinct
TCR complexes present across the entire naive pool.  The diversity of
T-cell clones, or ``immunoclones'', characterized by the number of
distinct TCR complexes among the cell population, provides the extent
of antigen specificity.  Unique TCR complexes are generated during
T-cell development in the thymus, via recombination of genes encoding
the V and J domains of the TCR$\alpha$ chain and the V, J, and D
domains of the TCR$\beta$ chain, along with additional insertion and
deletion of nucleotide fragments~(\citealt{JANEWAY2012}).
Combinatorially, a possible $\Omega_0 \sim 10^{15}-10^{20}$ unique TCR
complexes may be assembled via this rearrangement
process~(\citealt{LAYDON2015}), but only $\Omega \sim (0.05) \times
\Omega_0$ of those rearrangements are functionally
viable~(\citealt{YATES2014}), as determined by positive and negative
selection tests in the thymus, which screen for appropriate reactivity
to self-peptide/MHC molecules.  Each TCR is activated by at least one
peptide fragment presented via MHC molecules on the surface of an
antigen-presenting cell, thus loss of naive TCR structural diversity
limits the number of new antigens to which the full naive T-cell pool
can respond.  Naive cells are also suspected to suffer major
functional deficiencies in aging, such as diminished binding affinity
and proliferative capacity after antigenic
stimulation~(\citealt{MORO-GARCIA2013}).  While these effects have
been studied mostly using murine models to date~(\citealt{APPAY2014}),
they are not yet well understood in humans and are beyond the scope of
this paper.

The total abundance of naive T-cells, which inhabit both blood and
lymphatic tissue, can be reliably estimated from measurements in small
samples~(\citealt{WESTERMANN1990,BAINS2009}).  Recently, Westera
\textit{et al.}~\citeyearpar{WESTERA2015} estimated an $\sim 52\%$
decrease in the naive T-cell population in aging.  In contrast,
accurate estimation of full-organism TCR structural diversity is
currently impeded by experimental imprecision and the inability to
extrapolate small sample data to the full
organism~(\citealt{LAYDON2015}).  Experimentation typically entails
DNA sequencing of the TCR$\alpha$ or--more commonly--$\beta$ chain, in
particular the complimentarity-determining region 3 (CDR3), which is
the site of TCR binding to antigenic peptide and most significant
basis for diversity~(\citealt{JANEWAY2012}).

Increasingly sophisticated deep sequencing methods have improved
estimates for the lower bound on TCR diversity but direct estimation
of TCR diversity remains a challenge due to various experimental
complications, such as the inability to detect rare clonotypes,
sequencing errors, and inaccurate measurement of clonotype frequencies
resulting from inconsistencies in polymerase chain reaction (PCR)
amplification~(\citealt{LAYDON2015}). Predicting full-organism TCR
diversity from a small sample is typically formulated as an ``unseen
species problem", and one of many canonical solutions to such a
problem is employed in conjunction with experimental
data~(\citealt{CHAO1984,CHAO1992,COLWELL1994}), but the true
relationship between sample and full diversity is fundamentally
elusive.

Despite variations across experimental measurements of TCR diversity,
its age-related loss has been consistently observed.  An early study
conducted by Naylor \textit{et al.}~\citeyearpar{NAYLOR2005} predicted
a TCR$\beta$ chain diversity of $\sim 2 \times 10^7$ that persisted in
donors through age $60$, before dropping by two orders of magnitude to
$\sim 2 \times 10^5$ at age $70$.  More recently, Britanova \textit{et
  al.}~\citeyearpar{BRITANOVA2014} collected samples from donors of
all ages and observed an approximately linear decrease in TCR$\beta$
CDR3 diversity from $\sim 7 \times 10^6$ in youth ($6-25$ years) to
$\sim 2.4 \times 10^6$ in advanced age ($61-66$ years).  Qi \textit{et
  al.}~\citeyearpar{QI2014} obtained a particularly high lower bound
estimate of $\sim 10^8$ unique TCR$\beta$ sequences in youth ($20-35$
years), which declined two- to five-fold in advanced age ($70-85$
years).

Note that only the TCR$\beta$ chain is sequenced in these experiments.
Sequencing of both the $\alpha$ and $\beta$ chains would potentially
produce a more accurate measure of TCR diversity, but the same
experimental limitations preclude complete analysis.  The measurement
of diversity is further complicated by the potentially large disparity
between structural diversity and ``functional diversity"--that is, the
number of antigens to which the T-cell pool is capable of responding.
Due to the potential for crossreactivity, in which one TCR might
respond to many structurally similar peptide fragments, it is possible
that actual TCR diversity is much higher than structural diversity
indicates.  It has been speculated that one TCR might respond to as
many as $10^6$ different peptide epitopes~(\citealt{MASON1998}).

To obtain lifetime estimates of TCR structural diversity, and develop
an informed context for discussion of functional diversity, we
introduce a mechanistic mathematical model of the generation and
replenishment of the lymphocyte pool from birth through the end of
life.  Although experimental assessments of full-system information
remain challenging, measurements for the dynamics of each component
related to the T-cell population can be found throughout the
literature. Our mathematical approach combines the knowledge of these
individual components to study their interplay, leading to an
understanding of the full-system dynamics.  By extending previous
model studies of total cell
counts~(\citealt{MEHR1996,MEHR1997,RIBEIRO2007,BAINS2009,BAINS20092,HAPUARACHCHI2013,MURRAY2003,REYNOLDS2013}),
our multi-component formulation is able to efficiently track the total
number of distinct T-cell clones, allowing for a full-system
assessment of TCR structural diversity.

\section{Mathematical Models and Results}
\label{sec:2}

We develop our mathematical model by first constructing the equation
governing the total population size of the naive T-cell pool in
Sec.~\ref{sec:2.1}, through which we quantitatively constrain the
primary parameters of our model using experimental measurements found
in previous literature. The model that describes the evolution of
immunoclones is derived in Sec.~\ref{sec:2.2}, allowing us to define
and estimate the diversity of the T-cell population in
Sec.~\ref{sec:2.3}.  In Sec.~\ref{sec:2.4}, we inspect the impact of
sampling on the estimate of immunoclone diversity, as in practice it
is only possible to extract a small fraction of the entire T-cell
population from a body.

\subsection{Total T-cell population model}
\label{sec:2.1}

There are three fundamental immunological mechanisms
that sustain the naive T-cell pool: 1) export of mature naive T-cells
from the thymus, 2) peripheral proliferation, and 3) cell removal from
the naive pool due to death or phenotypic changes. These basic
mechanisms constitute a birth-death-immigration process 
described by the ordinary differential equation,

\begin{align}
\frac{{\rm d}N(t)}{{\rm d}t} = \gamma(t) + p N(t) - \mu(N) N(t),
\label{eq:EQNN}
\end{align}
where $N(t)$ denotes the total T-cell count, $\gamma > 0$ denotes the
rate of thymic output, $p > 0$ denotes the rate of proliferation, and
$\mu (N) > 0$ denotes the rate of population-dependent regulated
cellular death or loss of naive phenotype.


While more complex feedback mechanisms have been proposed
~(\citealt{MEHR1997}), other experiments have shown that thymic export
is independent of naive T-cell
counts~(\citealt{RIBEIRO2007,BERZINS1998,METCALF1963}), it is
well-established that the export rate consistently decays throughout
the human lifespan~(\citealt{MURRAY2003}).  The lifelong decline of
thymic export is caused by thymic involution which leads to
degradation of structural integrity and functional capacity of the
thymus with age~(\citealt{STEINMANN1985}). The age dependence of the
thymic export rate of newly-trained T-cells is often approximated by
an exponentially decaying function, $\gamma(t) = \gamma_0 e^{-at}$,
where $\gamma_0 > 0$ is the maximum rate of thymic output that arises
in early years, and $a > 0$ is the rate of decrease in thymic output.

The immune systems of vertebrates maintain a healthy amount of naive
T-cells through complex homeostatic mechanisms, which include
controlled production and distribution of common gamma chain
cytokines, particularly IL-7, to the naive pool~(\citealt{FRY2005}).
IL-7 is secreted by stromal and endothelial cells in the thymus, bone
marrow, and lymphatic tissue, providing T-cells with necessary
survival signals.  In lymphoreplete conditions, competition for this
limited resource regulates population
size~(\citealt{BRADLEY2005,TAN2001,VIVIEN2001}), but in lymphopenic
conditions, high levels of IL-7 resulting from low T-cell counts can
even stimulate cellular proliferation.  While IL-7 concentration may
be explicitly formulated in a mathematical model of the peripheral
T-cell population, as in the work of Reynolds \textit{et
  al.}~\citeyearpar{REYNOLDS2013}, most models incorporate IL-7
regulation implicitly in the form of carrying capacity, assuming quick
equilibration in a state of competition for IL-7 in the presence of a
given number of T-cells. Such simplification commonly leads to the
dependence on total cell counts of both cell proliferation and cell
death rates, considering the cytokine's dual role under lymphoreplete
and lymphopenic conditions described above. Our model assumes
cell-count dependence only of the cell death rate, focusing on
scenarios of \textit{healthy} aging, i.e., lymphoreplete
conditions. We thus assume an $N$-dependent cell death rate
of the form 

\begin{equation}
\mu(N) = \mu_0 + {\mu_1 N^2\over N^2 + K^2},
\label{eq:MUN}
\end{equation}
where the first term, $\mu_0>0$, is the basal rate of cellular death.
The second one describes the IL-7-mediated regulation of cell
death, with $\mu_1 > 0$ representing the maximal increase to the death
rate as $N \to \infty$. 
%
%
The quantity $K$ is analogous to a ``carrying capacity'' and dictates
the population at which signalling induced death starts to limit the
population.  The constant rate of cellular proliferation under healthy
conditions is supported by recent studies of Westera \textit{et
  al.}~\citeyearpar{WESTERA2015}, showing nearly identical naive
proliferation rates at young and old ages during moderate age-related
non-lymphopenic loss of naive cells.  IL-7 induced proliferation can
arise in \textit{unhealthy} lymphopenic conditions typically found in
severe disease of the immune system~(\citealt{BRASS2014}), cytotoxic
drug use~(\citealt{GERGELY1999}), radiation
treatment~(\citealt{GROSSMAN2015}), or other abnormal
situations. These scenarios are, however, beyond the scope of our
analysis.

\begin{figure}[h]
    \includegraphics[width=\textwidth]{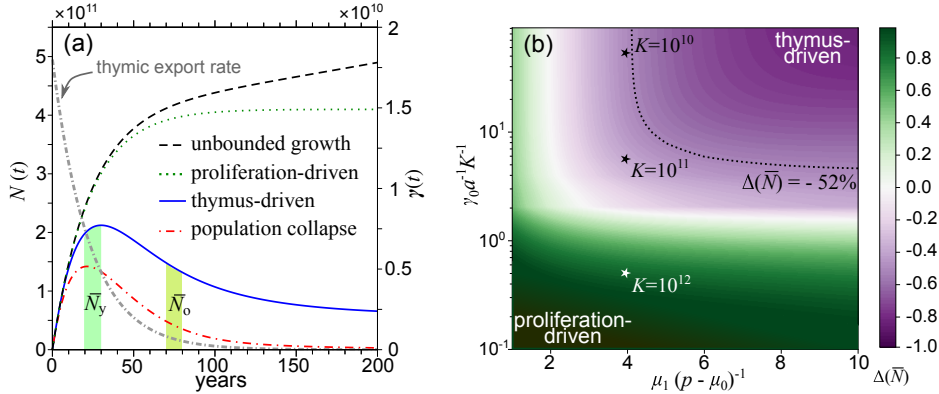}
    \captionsetup{name=Fig.}
    \caption{ {\bf Qualitative behavior of the total T-cell population
        model (Eqs.~\ref{eq:EQNN}, \ref{eq:MUN}).}  (a) The total
      T-cell population $N(t)$ as a function of time (in years) for
      four qualitatively distinct scenarios.  Unbounded growth arises
      when $\mu_1 \left( p - \mu_0 \right)^{-1} < 1$.  and the T-cell
      population collapses when $(p - \mu_0) a^{-1} < 0$. Outside of
      these two regimes, $N (t)$ converges asymptotically to a
      positive steady state as $\gamma (t) \to 0$.  If $\gamma_0
      a^{-1} K^{-1} \ll 1$, $N (t)$ is driven primarily by homeostatic
      proliferation and increases monotonically towards the constant
      plateau.  Increasing $\gamma_0 a^{-1} K^{-1}$ leads to a
      transition from proliferation-driven scenario to thymus-driven
      populations, in which $N(t)$ reaches a peak value before
      converging to the steady state.  The decaying thymic export rate
      $\gamma(t)$ is alongside of the $N(t)$ curves as a reference. To
      quantify the decrease in cell counts with age, we define
      $\bar{N}_{\rm y}$ as the average of $N(t)$ between ages $20$ and
      $30$, and $\bar{N}_{\rm o}$ between $70$ and $80$; then $\Delta
      \left( \bar{N} \right) = \left( \bar{N}_{\rm o} - \bar{N}_{\rm
        y} \right) / \bar{N}_{\rm y}$ is the relative change in cell
      counts.  The parameter values used are $\gamma_0 = 1.8 \times
      10^{10}$, $a = 0.044$, and $K = 10^{10}$ and $p = 0.022$, $\mu_0
      = 0.017$, and $\mu_1 = 0.004$ for unbounded growth, $p = 0.17$,
      $\mu_0 = 0.18$ and $\mu_1 = 0.04$ for the collapse scenario, $p
      = 0.18$, $\mu_0 = 0.17$, and $\mu_1 = 0.01001$ for the
      homeostasis-driven case, and $p = 0.18$, $\mu_0 = 0.17$, and
      $\mu_1 = 0.04$ for the thymus-driven case.  The initial value is
      $N(1) = 10^{11}$ at $t = 1$ year.  (b) $\Delta \left( \bar{N}
      \right)$ as a function of $\gamma_0 a^{-1} K^{-1}$ and $\mu_1
      \left( p - \mu_0 \right)^{-1}$. When $\gamma_0 a^{-1} K^{-1}$
      and $\mu_1 \left( p - \mu_0 \right)^{-1}$ are small, $N(t)$ is
      driven primarily by proliferation and keeps increasing well into
      old age, leading to positive $\Delta(\bar{N})$
      values. Conversely, for large $\gamma_0 a^{-1} K^{-1}$ and
      $\mu_1 \left( p - \mu_0 \right)^{-1}$, thymic export dominates
      and $N(t)$ peaks at early ages, resulting in negative
      $\Delta(\bar{N})$.  The black dotted curve corresponds to
      $\Delta(\bar{N}) = - 52\%$ as previously reported by Westera
      \textit{et al.} for human adults.  At fixed $\mu_1 \left( p -
      \mu_0 \right)^{-1} = 4$, we are able to reproduce this curve by
      setting $\gamma_0 a^{-1} K^{-1} \simeq 41$ (corresponding to $K
      = 10^{10}$ for our choice of parameter values).  The value of
      $\Delta(\bar{N})$ increases with decreasing $\gamma_0 a^{-1}
      K^{-1}$ and become positive when $\gamma_0 a^{-1} K^{-1}
      \lesssim 1$. Here, we fixed $(p - \mu_0) a^{-1} = 0.2$ and $a =
      0.044$.}
    \label{fig:QUALITATIVE}
\end{figure}

Our model has six adjustable parameters, $\gamma_0$, $a$, $p$,
$\mu_0$, $\mu_1$ and $K$. The first four are biologically inherent to
the mechanism of T-cell homeostasis, and have been measured
experimentally in humans and rodents.  The last two have to be
constrained via parameter sweeps to match relevant experimental
observations. Fig.~\ref{fig:QUALITATIVE}(a) illustrates four
qualitatively distinct evolution trajectories of $N(t)$ that may arise
from simulations of the model in the presence of a decaying thymic
export rate $\gamma (t)$ (gray dash-dotted curve).  
To non-dimensionalize Eqs.~\ref{eq:EQNN},~\ref{eq:MUN}, we use
$a^{-1}$ to rescale $t$ and $K$ to rescale $N$. The qualitative
behavior of our model is thus controlled by three independent
parameters: $\gamma_0 a^{-1} K^{-1}$, $(p - \mu_0) a^{-1}$, and $\mu_1
\left( p - \mu_0 \right)^{-1}$. The black dashed curve arises when
$\mu_1 \left( p - \mu_0 \right)^{-1} < 1$. In this case cell
proliferation always exceeds cell death, leading to unbounded
expansion of the naive T-cell population. This scenario is
unrealistic, except perhaps during a period of lymphopenia.  For
$\mu_1 \left( p - \mu_0 \right)^{-1} \ge 1$, cell death is able to
balance cell proliferation at a homeostatic carrying capacity $N =
N_{\rm ss} (\gamma = 0)$, defined by $\mu(N_{\rm ss} (\gamma = 0)) =
p$, as $\gamma \to 0$.  As illustrated by the green dotted curve,
$N(t)$ rises and asymptotically converges towards $N_{\rm ss} (\gamma
= 0)$ provided that $\gamma_0 a^{-1} K^{-1} \ll 1$. We refer to this
scenario as being in the ``proliferation-driven'' regime, given that
the cell population is driven to $N_{\rm ss} (\gamma = 0)$ primarily
by homeostatic proliferation. The model's behavior makes a transition
from proliferation-driven to ``thymus-driven'' if we increase
$\gamma_0 a^{-1} K^{-1}$. As shown by the blue solid curve, $N(t)$,
driven by increased thymic export, overshoots and approaches $N_{\rm
  ss} (\gamma = 0)$ from above as $\gamma(t) \to 0$ asymptotically.
Finally, the red dash-dotted curve arises when $(p - \mu_0) a^{-1} \le
0$. In this case cell death always exceeds cell proliferation as
$\gamma(t) \to 0$, and $N (t) \to N_{\rm ss} (\gamma = 0) = 0$.  As
stated earlier, in this paper we focus on scenarios of healthy aging
(lymphoreplete) conditions, which immediately rules out the scenarios
of unbounded growth (black dashed curve) and complete collapse of the
T-cell population (the red dot-dashed curve), effectively constraining
our parameters to physiologically reasonable values $\mu_1 \left( p -
\mu_0 \right)^{-1} \ge 1$ and $(p - \mu_0) a^{-1} > 0$.

We can further quantitatively calibrate the parameter values using
experimental measurements in the literature.  The constant peripheral
proliferation rate $p$ has been measured by Westera \textit{et
  al.}~\citeyearpar{WESTERA2015} as $0.05$\% $\text{day}^{-1}$, or
equivalently $p = 0.18$ $\text{year}^{-1}$.  The basal death rate
$\mu_0$ can be estimated from the lifespan of T-cells. Based on data
from Vrisekoop \textit{et al.}~\citeyearpar{VRISEKOOP2008}, De Boer
and Perelson~\citeyearpar{DEBOER2013} obtain an average naive
$\text{CD4}^+$ T-cell lifespan of $\sim 5$ years and an average naive
$\text{CD8}^+$ lifespan of $\sim 7.6$ years.  Given the normal
$\text{CD4}^+\text{:CD8}^+$ ratio of 2:1, the average combined naive
T-cell clearance rate is $\mu_0 = \frac{1}{5.9}$ $\text{year}^{-1}$ =
$0.17$ $\text{year}^{-1}$.  Thymic involution with age can be
quantified by measuring the decrease in thymic epithelial
volume~(\citealt{STEINMANN1986}), based on which Murray \textit{et
  al.}~\citeyearpar{MURRAY2003} showed that thymic output decreases by
an average of $4.3\%$ per year between ages $0$ and $100$, implying a
decay factor of $a = |\ln(0.957)| \simeq 0.044$.  The rate of thymic
export has recently been measured for young adults ($20-25$ years old)
at $\sim 1.6 \times 10^{7}$ trained cells daily, or equivalently $5.8
\times 10^{9}$ per year~(\citealt{WESTERA2015}).  Assuming that this
rate is $\gamma(t)$ at $t = 25$ years, we can back-calculate $\gamma_0
= (5.8 \times 10^{9}) \times \left(\frac{100}{33.3}\right) \approx
1.75 \times 10^{10} \text{cell exports}/\text{year}$. Note that
these values of $p$, $\mu_0$, and $a$ satisfy the constraint $\left( p
- \mu_0 \right) a^{-1} > 0$ that prevents the T-cell population from
completely collapsing.

While direct experimental measurements of $\mu_1$ and $K$ are not
available in the literature, further inspection of
Fig.~\ref{fig:QUALITATIVE}(a) reveals that $\mu_1$ and $K$ determine
whether thymic export or homeostatic proliferation dominates the
evolution of $N(t)$.  Through the dimensionless parameters, $\gamma_0
a^{-1} K^{-1}$ and $\mu_1 \left( p - \mu_0 \right)^{-1}$, the time at
which $N (t)$ peaks and how fast it declines from the peak vary with
changes to the values of $\mu_1$ and $K$.  Recently, Westera
\textit{et al.}~\citeyearpar{WESTERA2015} reported a $52\%$ decrease
in total naive T-cell counts between young human adults and elderly
individuals, which we can use to quantitatively constrain $\mu_1$ and
$K$.  Let us define individuals of an age between $t = 20$ and $30$
years as young adults, and those between $t = 70$ and $80$ as the
elderly. Assuming that interpersonal heterogeneity unrelated to age
averages out over large sample sizes in clinical data, we may evaluate
$\bar{N}_{\rm y} = \frac{1}{10} \int_{20}^{30} N(t) {\rm d}t$ and
$\bar{N}_{\rm o} = \frac{1}{10} \int_{70}^{80} N(t) {\rm d}t$ as the
average naive T-cell counts respectively for the young and the
elderly, as illustrated by the shaded areas under the
thymus-domination curve in Fig.~\ref{fig:QUALITATIVE}(a). The relative
change in the naive T-cell count between young and elderly adults can
thus be evaluated as

\begin{equation}
\Delta(\bar{N}) = {(\bar{N}_{\rm o} -
\bar{N}_{\rm y}) \over \bar{N}_{\rm y}}.
\end{equation}

\noindent Fig.~\ref{fig:QUALITATIVE}(b) plots $\Delta(\bar{N})$ as a
function of $\gamma_0 a^{-1} K^{-1}$ and $\mu_1 \left( p - \mu_0
\right)^{-1}$, with $a = 0.044 \textrm{ year}^{-1}$ for converting the
dimensionless time to years to compute $\bar{N}_{\rm y}$ and
$\bar{N}_{\rm o}$.  When $\gamma_0 a^{-1} K^{-1} \lesssim 1$ and
$\mu_1 \left( p - \mu_0 \right)^{-1} \lesssim 2$, $\Delta(\bar{N}) >
0$.  Note that the homeostatic carrying capacity when $\gamma (t) = 0$
is $N_{\rm ss} (\gamma = 0) = K \left( \mu_1 (p - \mu_0)^{-1} - 1 \right)^{-1}$.
 A small $\gamma_0 a^{-1} K^{-1}$ value represents a relatively low thymic
export rate, and the carrying capacity increases rapidly as $\mu_1
\left( p - \mu_0 \right)^{-1} \to 1$, both of which make it
challenging for thymic output to fill up the T-cell pool to carrying
capacity before $\gamma (t)$ considerably decays within $t \sim
a^{-1}$. As a result, $N(t)$ does not reach a peak value at a young
age and continues increasing into old age. The $\approx 52\%$ decrease
in naive T-cell counts reported by Westera \textit{et
  al.}~\citeyearpar{WESTERA2015} is depicted by the black dotted
curve. If we set $\mu_1 \left( p - \mu_0 \right)^{-1} = 4$, our model
can be calibrated to reproduce this decrease in the cell count by
choosing $K = 10^{10}$ ($\gamma_0 a^{-1} K^{-1} \simeq 41$ with
$\gamma_0 = 1.8 \times 10^{10}$ and $a = 0.044$). In contrast, $K =
10^{12}$ yields $\gamma_0 a^{-1} K^{-1} \simeq 0.41$, leading to an
increase in the cell count ($\Delta(\bar{N}) \simeq 0.63$).  In
between, $K = 10^{11}$ results in a moderate decrease in the cell
count ($\Delta(\bar{N}) \simeq -0.33$). For the rest of the paper, we
fix $K = 10^{10}$ and $\mu_1 \left( p - \mu_0 \right)^{-1} = 4$, or
equivalently $\mu_1 = 0.04$ given that $p = 0.18$ and $\mu_0 = 0.17$,
so that the age-related decline of $N(t)$ in our model is consistent
with Westera \textit{et al.}~\citeyearpar{WESTERA2015}.

\begin{figure}[h]
  \centering
      \includegraphics[width=0.8\textwidth]{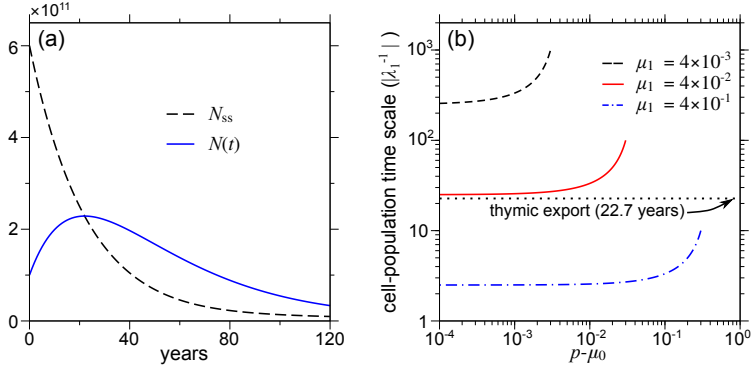}
            \captionsetup{name=Fig.}
            \caption{{\bf Comparison of Thymic Export and
                Cell-Population Evolution Time Scales.}  (a) Plots of
              $N(t)$ and $N_{\rm ss}$ show discrepancy.  The
              $\gamma(t)$-dependence makes $N_{\rm ss}$ decline
              monotonically with the exponentially decaying thymic
              export, and $N_{\rm ss}$ approaches a small positive
              value as $\gamma(t) \to 0$.  The solution $N(t)$ evolves
              towards $N_{\rm ss}$ but never catches up with it
              because of a slower evolution time scale.  (b)
              Comparison of timescales of thymic atrophy and
              cell-population evolution.  Thymic atrophy is the faster
              mechanism for most choices of the system's parameters.
              Increasing $\mu_1$ shortens the time scale of clone
              evolution, indicating that the steady state solution can
              be a reasonable approximation to the fully
              time-dependent solution at very large $\mu_1$ and very
              small $p - \mu_0$.  Here, varying $N_{ss}$ within the
              range $[10^{10}, 10^{12}]$ yields almost identical
              results, and the values of $\gamma_0$ and $K$, chosen
              within the reasonable parameter regime, do not affect
              the results significantly.  Parameter values used are
              $\gamma_0 = 1.8 \times 10^{10}$, $a = 0.044$, $p =
              0.18$, $\mu_0 = 0.17$, $K = 10^{10}$, $\Omega =
              10^{16}$.  For (a) $\mu_1 = 0.04$, and the initial
              condition is $N(1) = 10^{11}$.}
    \label{fig:ERROR}
\end{figure}

Note that there exist two intrinsic timescales in Eq.~\ref{eq:EQNN};
thymic export decays at a rate $a$, while the homeostatic time scale
is controlled by $p$, $\mu_0$, and $\mu_1$.  If homeostasis is much
faster than thymic involution, the solution of $N(t)$ will quickly
converge to the quasisteady state solution as $\gamma(t)$ evolves. We
compare these two solutions in Fig.~\ref{fig:ERROR}(a), where the
quasisteady-state solution is obtained by solving for the steady-state
solution $N_{\rm ss}$ of Eq.~\ref{eq:EQNN} with fixed $\gamma(t)$ at
each time $t$, and $N_{\rm ss} (\gamma(t))$ (black dashed curve)
decreases monotonically with age due to the continuous decline of
$\gamma(t)$. In contrast, $N(t)$ (blue solid curve) slowly rises from
the initial conditions $N(1) = 10^{11}$ and does not approach the
quasisteady-state level until age $\approx 20$ years.  The trajectory
of $N(t)$ then overshoots the declining $N_{\rm ss} (\gamma(t))$,
reaches a peak value, and reverts course to go after $N_{\rm ss}
(\gamma(t))$.  However, $N(t)$ never catches up with $N_{\rm ss}
(\gamma(t))$ before the latter reaches a steady state of very low cell
counts.  That $N(t)$ keeps lagging behind $N_{\rm ss} (\gamma(t))$
indicates that the timescale for the full model solution to converge
to the steady state is slower than the evolution of the nonautonomous
term $\gamma(t)$. The results here suggest that steady-state solutions
cannot adequately describe the temporal evolution of the T-cell
population in the biologically relevant range of parameter values that
we have implemented.  It is necessary to numerically compute the
time-dependent solutions for the full nonautonomous equation.

Indeed, we find a disparity in the rates at which thymic export decays
and the steady state solutions evolve.  The latter is provided by the
inverse of the eigenvalue of Eq.~\ref{eq:EQNN} linearized around $N =
N_{\rm ss} (\gamma(t))$.  The eigenvalue takes the form $\lambda_1 =
p_0 - (\mu_0 + \mu_1((3 N_{\rm ss}^2 K^2 + N_{\rm ss}^4)/((K^2+ N_{\rm
  ss}^2)^2))$.  Simulations in Fig.~\ref{fig:ERROR}(b) show that for
the biologically relevant parameter values we have implemented, the
cell-population evolution timescale, $|\lambda_1|^{-1}$ (red solid
curve), is generally longer than the timescale of thymic involution
($a^{-1} \simeq 22.7$ years for $a = 0.044$ as denoted by the
horizontal black dotted line).  Hence the nonautonomous solutions
$N(t)$ are expected to lag behind the thymus-driven steady-state
solutions $N_{\rm ss}$.  For $N (t)$ to be reasonably approximated by
$N_{\rm ss}$, the cell population has to evolve much faster than
thymic involution, corresponding to the regime of very large $\mu_1$,
as indicated by the blue dash-dotted curve, where cell death is
extremely sensitive to the cell population size.

\subsection{Clonotype Abundance Distributions}
\label{sec:2.2}
Quantification of the populations of individual clonotypes would
require analysis of models that track the population dynamics of naive
T-cells of each TCR type. Assuming the same population dynamics for
each T-cell clonotype $i$, which may be appropriate for certain
scenarios, the evolution of the expected cell count $n_i(t)$ may be
deduced from Eq.~\ref{eq:EQNN} and take the following generalized
form,

\begin{align} \label{eq:NIEQ}
\frac{{\rm d}n_i}{{\rm d}t} = \frac{\gamma(t)}{\Omega} + p n_{i}-\mu(N)n_{i},
\end{align}

\noindent where $\gamma(t)/\Omega$ represents thymic export of naive
T-cells of \textit{each} clonotype (the total thymic export rate
normalized by the total number of viable TCR combinations $\Omega$),
and $N(t) = \sum_i n_i(t)$.  Within the framework of these ``neutral''
models, basic qualitative behaviors of T-cell population dynamics have
been investigated, particularly for scale-invariant properties that
can be studied in a reduced system~(\citealt{LYTHE2016,DESPONDS2015}).
Indeed, the total numbers of T-cell clonotypes $\Omega$ in rodent or
human bodies are prohibitively large for direct numerical simulations
of the full system using Eq.~\ref{eq:NIEQ}.  It is thus common to
reduce the full system to a more manageable size with the assumption
that the phenomena under investigation are scale-invariant. However, it
is sometimes difficult to assert whether a certain property really
does not change in a re-scaled system, as nonlinear phenomena, such as
the Allee effects, often arise in population dynamics and cast doubt
on the scalability of the system. Moreover, some properties, such as
the thymic export rate $\gamma(t)$, are naturally scale dependent. It
is not always clear how these quantities should be re-scaled in a
reduced system, and they have usually been omitted by simplification
arguments in previous models, which limits the applicability of these
models.

In particular, thymic involution is known to be associated with the
age-related loss of T-cell diversity. Without the explicit inclusion
of the thymic export rate, such loss of T-cell diversity cannot be
properly investigated.  To facilitate a more manageable full-system
model, we consider a formulation that tracks how the expected number
of clones of a given size changes with time. By focusing on clone
count rather than the explicit cell count of each distinct clonotype,
we are able to effectively reduce the number of tracked variables and
thus the dimension of the model.  This representation was used by
Ewens in population genetics~\citeyearpar{EWENS1972}, by Goyal
\textit{et al.}~\citeyearpar{GOYAL2015} in the context of
hematopoietic stem cell population dynamics, and by Desponds
\textit{et al.} in the context of T-cells~\citeyearpar{DESPONDS2017}.
We define $\hat{c}_k(t)$ to be the number of clones represented by
exactly $k$ naive T-cells in the organism at time $t$:

\begin{align} \label{CKDEF}
\hat{c}_k(t) = \sum_{i = 1}^{\Omega} \delta_{n_i(t),k},
\end{align}
where the Kronecker delta function $\delta_{x,y} = 1$ when $x = y$ and
$0$ otherwise.  By lumping clonotypes of the same cell count into one
single variable $\hat{c}_k$, this alternative formulation can
efficiently describe changes to the TCR clone diversity in the full
system, albeit at the expense of the ability to distinguish each
specific clonotype~(\citealt{MORRIS2014,MORA2016}).  Individual clone
information is lost, and $n_i(t)$ cannot be recovered from
$\hat{c}_k(t)$ after the transformation in Eq.~\ref{CKDEF}.
Nonetheless, the amount of computation can be significantly reduced by
truncating $\hat{c}_k(t)$ at a reasonably large $k$, as few large
clones exist in realistic scenarios, and $\hat{c}_k(t)$ for large $k$
is negligible.  Letting $c_0(t)\equiv \langle \hat{c}_{0}(t)\rangle$
denote the expected number of all possible (thymus-allowed) clonotypes
unrepresented in the periphery at time $t$, and $c_k(t)\equiv \langle
\hat{c}_{k}(t)\rangle$ the expected number of clones of size $k$ at
time $t$, a closed set of equations governing the evolution of
$c_{k}(t)$ can be derived from Eq.~\ref{eq:NIEQ} in the mean-field
limit,

\begin{align}
\frac{{\rm d}c_k(t)}{{\rm d}t} &= \frac{\gamma(t)}{\Omega} \left[c_{k-1} - c_k\right]
+ p\left[(k-1)c_{k-1} - k c_k\right] + \mu(N) \left[(k+1)c_{k+1} - k c_k\right], \label{eq:CLONEEQ2}
\end{align}
where $N(t) = \sum_{i}^{\infty} n_{i}(t) = \sum_{\ell=1}^{\infty}\ell
c_{\ell}(t)$.  The expected values $c_{k}(t)$ are also called species
abundances in the ecology literature.  The number of unrepresented
clones is $c_0 = \Omega - \sum_{k=1}^\infty c_k$, and summing
Eq.~\ref{eq:CLONEEQ2} multiplied by $k$ over $k = 1, 2, \cdots $
recovers Eq.~\ref{eq:EQNN}.  The mean-field assumption is articulated
in terms such as $\mu(\sum_{\ell} \ell \hat{c}_{\ell}) \hat{c}_{k}$
that involve higher-order products of $\hat{c}_{k}$ rather than
correlations of products of $\hat{c}_{k}$.

We have found (unpublished) that this mean-field approximation breaks
down only when $\gamma/\mu < 1/\Omega \ll 1$ for which the total
population is proliferation driven and the quasistatic configuration
is $N \sim K$ and all $c_{k} \sim 0$ except $c_{N}$. Thus, we 
reasonably assume that $\gamma(t) > \mu/\Omega$
allowing the use of the mean-field equations \ref{eq:CLONEEQ2}.

In Eq.~\ref{eq:CLONEEQ2}, the terms in the forms
of $(\gamma(t)/\Omega)c_k$, $p k c_k$, and $\mu(N) k c_k$ respectively
represent the effect of thymic export, homeostatic proliferation and
cell death on a T-cell clone already represented by $k$ cells in the
peripheral blood. Adding one cell via thymic export or homeostatic
proliferation moves one clone from the $c_k$-compartment to the
$c_{k+1}$-compartment, while the death of one cell shifts one clone
from the $c_k$-compartment to the $c_{k-1}$-compartment.  We
approximate the proliferation rate $p$ as a constant, at which rate
all cells of all clones of size $k$ replicate via homeostatic
proliferation.  Proliferation reduces $c_k$ and increases
$c_{k+1}$. Terms of the form $\mu(N) k c_k$, where the IL-7 regulated
death rate $\mu(N)$ is given by Eq.~\ref{eq:MUN}, reduce $c_k$ and
increase $c_{k-1}$.

For a healthy aging adult, the TCR repertoire is mostly comprised of
small clones with the probability of finding large clones decreasing
with clone size $k$.  To numerically solve
Eq.~\ref{eq:CLONEEQ2}, we thus truncate the model
at a maximum clone size $M \gg 1$, beyond which the probability of
finding a clone is assumed negligible.  For our implementation of the
truncation, please see Appendix~\ref{sec:APP1}.  In Fig.~\ref{fig:CKVSM}(a) we
examine the effect of the truncation clone size $M$, showing
sufficient convergence of $c_{10}$ at $t = 40$ and $70$ to fixed
values when $M \gtrsim 30$, which indicates that further inclusion of
clones beyond $c_{30}$ has little effects on the solution for $t
\lesssim 70$ years. For numerical simulations of
Eq.~\ref{eq:CLONEEQ2} in this paper, we set $M =
200$ to ensure minimal truncation errors.

\begin{figure}[h]
  \centering
    \includegraphics[width=\textwidth]{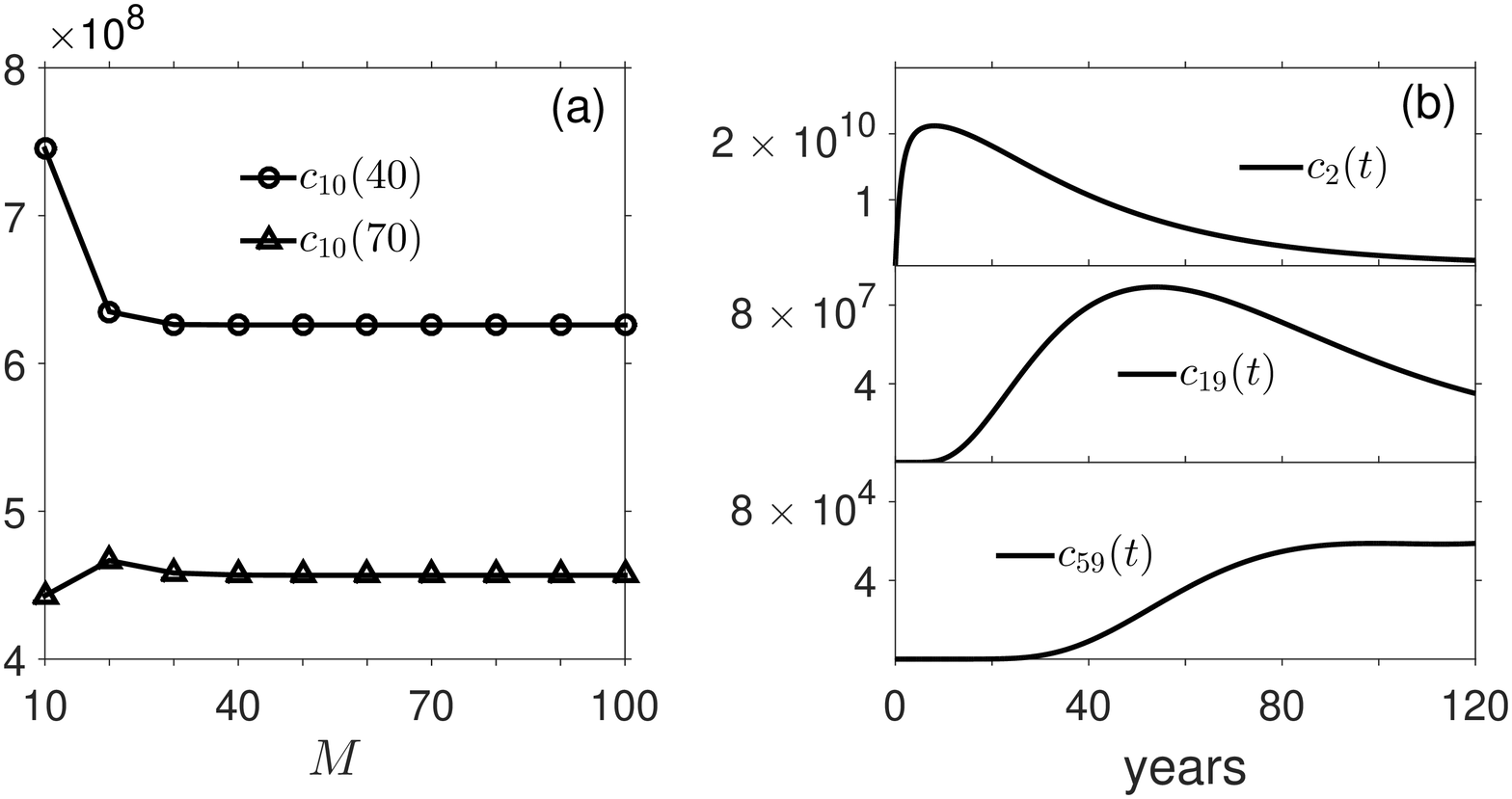}
    \captionsetup{name=Fig.}
    \caption{{\bf Simulations of
        Eq.~\ref{eq:CLONEEQ2}.}  (a) Effect of
      numerical truncation. We plot $c_{10}(40)$ and $c_{10}(70)$ as
      functions of $M$ for $10 \leq M \leq 100$.  Compartment sizes
      are effectively fixed when $M \gtrsim 30$.  (b) Temporal
      evolution of $c_k (t)$. We plot $c_2 (t)$, $c_{19} (t)$, and
      $c_{59} (t)$. Each $c_k (t)$ curve rises to a peak value
      and subsequently decreases. As $k$ increases, $c_k (t)$
      decreases in magnitude, and the time at which it reaches the peak value
      is pushed back. Parameter values: $\gamma_0 = 1.8 \times
      10^{10}$, $a = 0.044$, $p = 0.18$, $\mu_0 = 0.17$, $\mu_1 =
      0.04$, $K = 10^{10}$, $\Omega = 10^{16}$.  Initial values
      $c_1(1) = 10^{11}$, $c_0(1) = \Omega - 10^{11}$, $c_k(1) = 0$
      for all $k \geq 2$}
    \label{fig:CKVSM}
\end{figure}

Fig.~\ref{fig:CKVSM}(b) shows the temporal evolution of $c_k (t)$ for
$k = 2$, $19$, and $59$. As $k$ increases, the overall magnitude of
the $c_k (t)$ curve decreases, and the age at which $c_k (t)$ peaks
increases. For example, $c_2 (t)$ peaks around $t \lesssim 20$ years,
and there are many fewer clones of exactly two copies at old ages than
at young ages. In contrast, $c_{19}(t)$ peaks around age $55$, and the
numbers of clones that have exactly $19$ copies are roughly the same
between old and young ages, whereas the number of clones that have
exactly $59$ copies ($c_{59} (t)$) keeps increasing into old
ages.

The relatively earlier decline of $c_{k}(t)$ with smaller $k$ is
expected, considering that rare clones are introduced into the peripheral
circulation primarily by the thymus, which started to involute
after birth. With increasing $k$, the influence of thymic export on
$c_{k}(t)$ decreases, whereas the dependence on homeostatic
proliferation increases. Recalling that the rate of thymic
involution is faster than the time scale for homeostasis to drive
the clonal population towards equilibrium, the fast decline of the
rare clone population leaves room for larger clones to expand.

To accompany the steady state $N_{\rm ss}$, we compute analogous
fixed-$\gamma_0$ steady state values of the full system, $c_k^{\rm
  ss}$, in Appendix~\ref{sec:APP2}.  The steady states satisfy
$c_k^{\rm ss} \to 0$ as $\gamma_0 \to 0$ for all $1 \leq k \leq M$.
We further show that in spite of the fact that $c_k^{\rm ss} \to 0$,
Eq.~\ref{eq:CLONEEQ2} asymptotically yields a positive total cell
count $N = \lim_{M \to \infty} \sum_{k=1}^M k c_k^{\rm ss} > 0$ as $M
\to \infty$, qualitatively consistent with Eq.~\ref{eq:EQNN}.
Moreover, we prove in Appendix~\ref{sec:APP3} that solutions $c_{k}(t)$
of the full nonautonomous system satisfy $c_{k}(t) \to 0$ for all $k
\leq M$, with arbitrarily large $M$, as $t \to \infty$. This result is
completely independent of the assumed functional forms of the
proliferation and death rates, suggesting that manipulation of
homeostatic regulatory mechanisms cannot prevent the extinction of
small T-cell clones caused by decaying $\gamma(t)$.  We thus conclude
that thymic involution dictates the age-related decline of the TCR
diversity of the naive compartment.

\subsection{Diversity of the Naive T-cell Repertoire}
\label{sec:2.3}
By computing the functions $c_k$ that track the number of clones
consisting of $k$ cells, we should have sufficient information to
evaluate the variation in TCR structural diversity over a lifetime.
Expected TCR structural diversity or ``richness'' is the total number
of distinct clones present in the immune compartment, for which we
define a threshold TCR richness diversity,

\begin{align} \label{eq:RICHDEF}
R_q(t) = \sum_{k \geq q} c_k(t),
\end{align}

\noindent where $q \in \mathds{N}$ is a lower threshold, so that the
quantity $R_q(t)$ represents the number of clones of size at least $q$
present in the immune compartment at time $t$.  Situations in which such a
$q$-dependent threshold arise may include consideration of immune
surveillance, in which small clones may evade detection.

\begin{figure}[h]
    \centering
    \includegraphics[width=\textwidth]{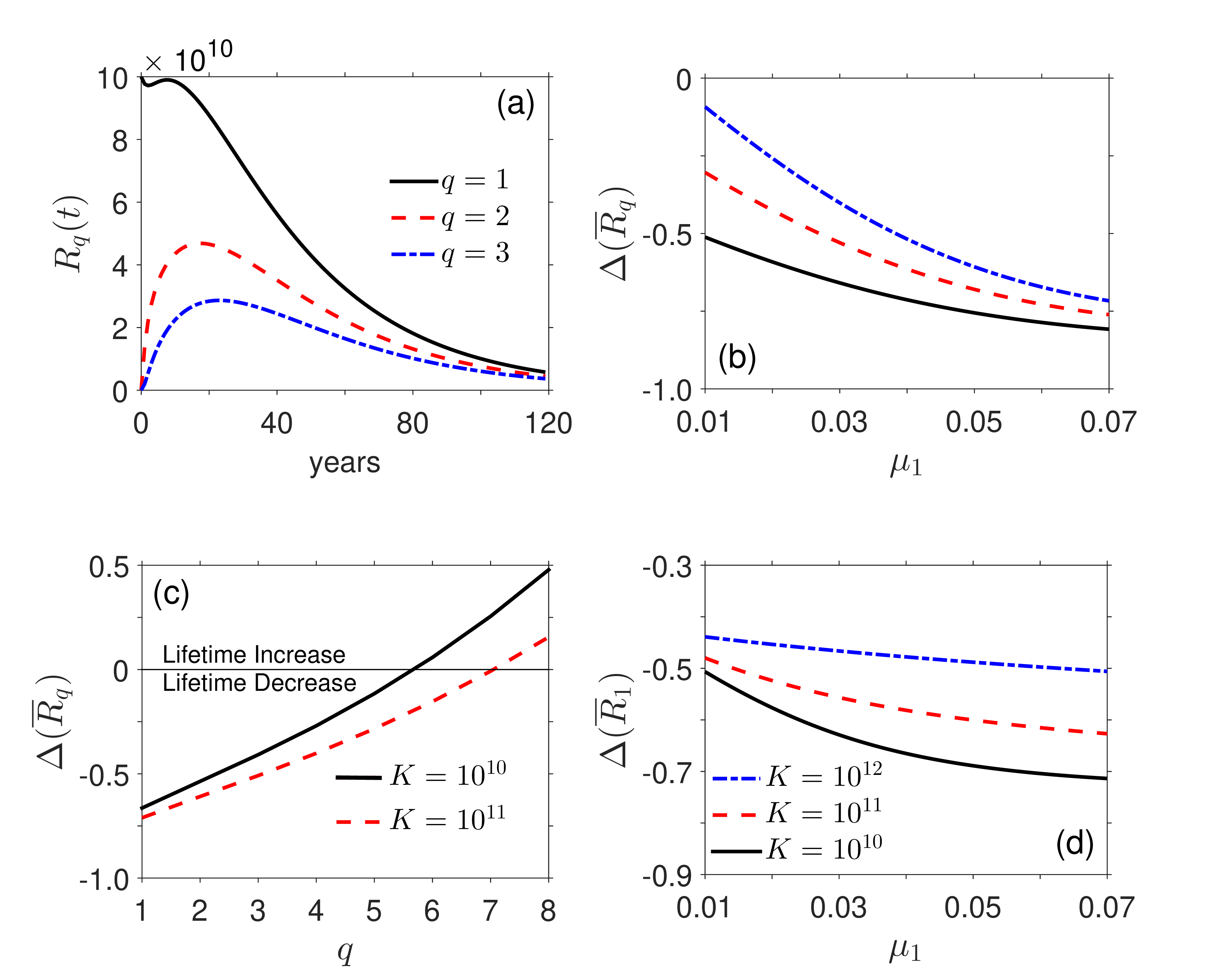}
    \captionsetup{name=Fig.}
    \caption{\textbf{Simulation of Threshold Richness Diversity.} (a)
      $R_q(t)$ as a function of $t$, for $q =$ 1, 2, 3.  $R_q$
      peaks at later times as $q$ increases. (b)
      $\Delta(\bar{R}_q(t))$ for varying $q$, $\mu_1$.  Higher
      $\mu_1$ correspond to more severe loss of T-cell clones in
      advanced age.  (c) $\Delta(\bar{R}_q)$ for varying $q$,
      $K$. Small values of $q$ result in a lifetime decrease to $R_q$,
      but larger values result in a lifetime increase.  This is due to
      the fact that $R_q$ peaks at later times as $q$ increases.  (d)
      $\Delta(\bar{R}_1)$ for varying $\mu_1$, $K$.  Initial
      values $c_0(1) = \Omega - 10^{11}$, $c_1(1) = 10^{11}$ $c_k(1) =
      0$ for $k \geq 2$.  Parameter values, when not varying: $\Omega
      = 10^{16}$, $K = 10^{10}$, $p_0 = 0.18$, $\mu_0 = 0.17$, $\mu_1
      = 0.04$, $a = 0.044$, $\gamma_0 = 1.8 \times 10^{10}$.}
    \label{fig:RICHNESS}
\end{figure}

As shown in Fig.~\ref{fig:RICHNESS}(a), $R_q (t)$ increases at young
ages, peaks at a mature age, and declines afterwards. For our previous
parameter values, the peak age of $R_1(t)$ is approximately $t \sim
16$.  Higher $q$ lead to older peak ages of $R_q(t)$,
consistent with the results in Fig.~\ref{fig:CKVSM}(b), in which
the number of larger clones peaks at older ages.

To compare $R_q (t)$ between the elderly
and young, we adopt the same criterion as with total cell counts
and compute window-averaged values of $R_q(t)$ between ages $20$ and
$30$ for the young and between ages $70$ and $80$ for the elderly.  By
defining $\bar{R}_{\rm y}(q) \equiv \frac{1}{10} \int_{20}^{30} R_q(t)
{\rm d} t$, $\bar{R}_{\rm o}(q) \equiv \frac{1}{10} \int_{70}^{80}
R_q(t) {\rm d} t$, we quantify the loss of richness by computing its
relative change.

\begin{equation}
\Delta(\bar{R}_{q}) \equiv {(\bar{R}_{\rm o}(q) - \bar{R}_{\rm
  y}(q))\over \bar{R}_{\rm y}(q)}.
\end{equation}
Using the same parameter values as in Fig.~\ref{fig:RICHNESS}(a), we
plot $\Delta(\bar{R}_{q})$ with respect to $\mu_1$ and $q$ in
Fig.~\ref{fig:RICHNESS} (b),(c).  In Fig.~\ref{fig:RICHNESS}(b),
$\Delta(\bar{R}_{q})$ decreases monotonically with increasing $\mu_1$,
suggesting that upregulated death rate exacerbates the age-related
loss of richness, and the impact is more significant for larger $q$.
Fig.~\ref{fig:RICHNESS}(c) shows that when $K=10^{10}$,
$\Delta(\bar{R}_{q}) < 0$ for $q \leq 4$.  This decreasing trend of
$R_{q}$ generally agrees with the loss of diversity observed in recent
experiments where measurements were available across multiple
ages~(\citealt{QI2014,BRITANOVA2014}).  For $q = 5, 6$,
$\Delta(\bar{R}_{q}) \approx 0$, and $R_{q}$ is nearly unchanged
between youth and advanced age. For $q \geq 7$, $\Delta(\bar{R}_{q}) >
0$, indicating higher $R_{q}$ at older ages.  Generally, the lifetime
decrease in $R_q(t)$ occurs with small $q$, whereas for large $q$, the
trend is reversed, in agreement with our discussion of
Fig.~\ref{fig:CKVSM}(b) and Fig.~\ref{fig:RICHNESS}(a) regarding peak
ages.  This phenomenon indicates that loss of diversity is primarily
due to the extinction of rare clones, which is consistent with the
observation made by Naylor \textit{et al.}~\citeyearpar{NAYLOR2005}.
In contrast, the number of larger clones increases over time, leading
to the lifetime increase to $R_q(t)$ at higher $q$.

Recent TCR-$\beta$ sequencing studies have attempted to estimate the
change in the repertoire richness of the naive T-cells with
age. Despite the difference in orders of magnitude regarding the total
number of circulated naive T-cell clones, these studies agreed
quantitatively in the ratio of the age-related loss of richness. For
example, Britanova \textit{et al.}~\citeyearpar{BRITANOVA2014}
estimated $\sim 7 \times 10^6$ clonotypes in youth (ages $6-25$), and
$\sim 2.4 \times 10^6$ in aged individuals (ages $61-66$), a roughly
$66$\% drop from the youth figure.  Similar measurements were also
reported by Qi \textit{et al.}~\citeyearpar{QI2014}, in which a
two-to-five-fold decline (i.e., a $50$\% -- $80$\% drop) between youth
(ages $20-35$) and advanced age (ages $70-84$) was observed.  These
results are quantitatively consistent with our computation of
$\Delta(\bar{R}_1)$ for $K = 10^{10}$ -- $10^{11.5}$ and $0.03 \leq
\mu_1 \leq 0.05$ in Fig.~\ref{fig:RICHNESS}(d), whereas the decline of
$R_{q}$ for $q \ge 2$ is not as pronounced as in these experimental
observations.

Also note that the loss of clonal richness is more severe than the
decrease in the total cell count between young and aged individuals.
In Fig.~\ref{fig:RICHNESS}(a) $\Delta(\bar{R}_{1})$ changes between
$\sim - 66\%$ and $\sim - 76\%$ for $0.03 \leq \mu_1 \leq 0.05$ and $K
= 10^{10}$.  In contrast, Fig.~\ref{fig:QUALITATIVE}(b) shows that for
the same parameter range, $\Delta(\bar{N})$ varies from $\sim - 30\%$
to $\sim - 62\%$.  However, the figures also reveal that richness is
relatively less sensitive to changes to the cellular death rate,
compared to the total cell count.  This outcome reflects the fact that
homeostatic cellular death is uniformly random across the entire naive
T-cell population.  The drop in richness is due to cell death within
small clones that drives these clones to extinction, as observed by
Naylor \textit{et al.}~\citeyearpar{NAYLOR2005}.  Increases to the
cellular death rate do not cause as much additional clonal extinction
as they do additional cellular extinction, as many surviving clones
are too large to wipe out by the death of a few cells.

\subsection{Sampling Statistics}
\label{sec:2.4}
Considering that naive T-cell richness is often assessed via small
blood samples, let us next use the same framework to examine the
relation between the detected clone sizes in small samples and the
true clone sizes in the full organism. As before, denote by $N$ the
total number of naive T-cells in the human's immune compartment, and
$Y \leq N$ the number of cells collected during sampling from among
the $N$ total.  We assume that the $N$ total cells consist of $R$
distinct clones, which we number from $1$ to $R$.  In this section, we
denote by $c_k^N$ the mean number of clones of size $k$ from among the
$N$ total cells in the full organism (denoted by $c_k$ in the previous
simulations), and by $c_k^Y$ the mean number of clones of size $k$ in
the sampling of $Y$ cells taken from the $N$ total cells.  Then the
expectation of $c_k^Y$, denoted by $\mathbb{E}[c_k^Y]$, is,

\begin{align} \label{eq:DEFEXPECTATION}
\mathbb{E}[c_k^Y] = \sum_{j=1}^R j P\left(c_k^Y = j\right),
\end{align}

\noindent where $P\left(c_k^Y = j\right)$ represents the probability
that there are precisely $j$ clones of size $k$ in the sampling.  Then
$\mathbb{E}[c_k^Y]$ may be expressed explicitly in terms of the
$c_k^N$ as:

\begin{align} \label{eq:FINALEXPECTATION}
\mathbb{E}[c_k^Y] &= \sum_{l=k}^R \frac{1}{\binom{N}{Y}} c_l^N
\binom{l}{k} \binom{N-l}{Y-k}.
\end{align}

\noindent (See Appendix~\ref{sec:APP4} for the detailed proof.)
The collection of
expressions given by Eq.~\ref{eq:FINALEXPECTATION} for $k = 1, 2,
\cdots, R$, yields a linear system of equations solvable for $c_k^N$,
using sampled data for the quantities $\mathbb{E}[c_k^Y]$.  More
specifically, if we define the vectors $\mathbf{\widehat{E}} :=
(\mathbb{E}[c_1^Y],\mathbb{E}[c_2^Y],\cdots,\mathbb{E}[c_R^Y],)$ and
$\mathbf{E} := (c_1^N,c_2^N,\cdots,c_R^N)$,
Eq.~\ref{eq:FINALEXPECTATION} can be written as $\mathbf{\widehat{E}}
= \mathbf{A E}$, where $\mathbf{A}$ is a constant matrix that has
non-zero elements only in the upper triangle, with non-zero diagonal
entry $\frac{1}{\binom{N}{Y}} \binom{N-k}{Y-k}$ in position $(k,k)$.
The equation can always be solved uniquely for $\mathbf{E}$ given
$\mathbf{\widehat{E}}$.  Thus the full size distribution $\mathbf{E}$
can be uniquely reconstructed from the expected mean sample size
distribution $\mathbf{\widehat{E}}$ measured experimentally, provided
that the latter can be reliably estimated through a sufficient number
of repeated samplings.

\begin{figure}[h]
    \centering
    \includegraphics[width=\textwidth]{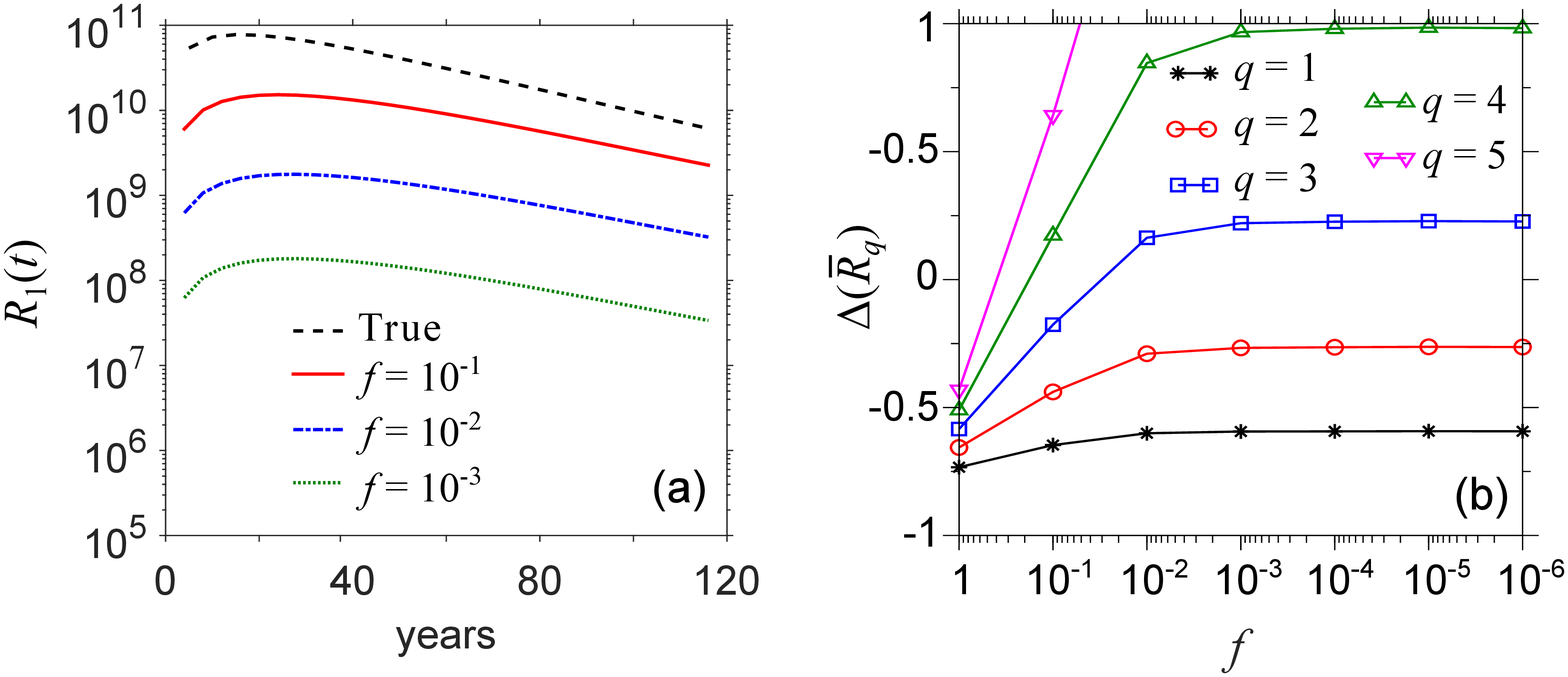}
    \captionsetup{name=Fig.}
    \caption{{\bf Comparison of Actual and Sampled Richness.}  (a)
      True lifetime $R_1$, as well as the expected $R_1$ that result
      from extracting 10$\%$, 1$\%$, and 0.1$\%$ of the total cell
      count for sampling.  ($Y = f \times N$, with $f = 10^{-1}$,
      $10^{-2}$, $10^{-3}$.) Each decrease to the sample size by one
      order of magnitude results in a decrease to the expected $R_1$
      by approximately one order of magnitude.  (b) The ratio of
      age-related TCR richness decline $\Delta \left( \bar{R}_q
      \right)$ as a function of sampling fraction $f$ for clone size
      thresholds $q = 1$ -- $5$.  As $f$ decreases, the value of
      $\Delta \left( \bar{R}_q \right)$ increases, indicating a
      lower estimate of the TCR richness decline.  When $f$ is very
      small, $\Delta \left( \bar{R}_q \right)$ becomes insensitive to
      further decreases to $f$.  Parameter values used: $\gamma_0 = 1.8
      \times 10^{10}$, $a = 0.044$, $p = 0.18$, $\mu_0 = 0.17$, $K_0 =
      10^{10}$, $\Omega = 10^{16}$, $\mu_1 = 0.04$.  Initial values
      $c_0(0) = \Omega$, $c_k(0) = 0$ for $k \geq 1$}
    \label{fig:SAMPLEvsK}
\end{figure}

In Fig.~\ref{fig:SAMPLEvsK}(a), we use Eq.~\ref{eq:FINALEXPECTATION}
to compute $\mathbb{E}[c_k^Y]$ from simulated $c_k^N$, comparing the
predicted sampling results for varying choices of $Y$.  The results
indicate that each decrease by one order of magnitude to the sample
size results in a decrease by roughly the same order of magnitude to
the predicted diversity.  Thus, diversity predictions vary with sample
size, and small samples do not result in accurate measurements of
diversity.

In Fig.~\ref{fig:SAMPLEvsK}(b) we examine how sampling may affect the
diagnosis of the age-related TCR richness decline $\Delta \left(
\bar{R}_q \right)$ defined in the previous subsection. We find $\Delta
\left( \bar{R}_q \right)$, which is negative, increasing with
decreasing sampling fraction $f$, revealing that sampling causes an
underestimate of the richness decline.  As previously discussed, the
decline of TCR richness at old ages is primarily due to the extinction
of small clones. Since small clones often evade detection during
sampling, their extinction is largely unaccounted for, leading to
lessened reduction of the richness measure. When $f$ is very small,
most of the small clones have escaped detection; thus decreasing $f$
further does not change $\Delta \left( \bar{R}_q \right)$.  Moreover,
we note that $\Delta \left( \bar{R}_1 \right)$, which is the most
straightforward measure for age-related loss of TCR richness, changes
from $-73\%$ for the full sample, to $-59\%$ for a sampling fraction
$f \le 10^{-3}$, which is close to the value of
$\Delta \left(\bar{R}_3 \right)$ for the full sample.  This reaffirms
our discussion in the previous subsection that a threshold $q > 1$ may
arise during the process of sampling.  The results here indicate that
when only a small fraction of a T-cell population is used to measure
$\Delta \left( \bar{R}_1 \right)$, clones fewer then three copies
largely evade detection, yielding a result equivalent to $\Delta
\left( \bar{R}_3 \right)$ of the full sample, which underestimates the
actual decrease of the TCR richness.

\section{Discussion}
\label{sec:3}
We have formulated a model of lifetime human naive T-cell population
dynamics, which traces T-cell lineages on the level of individual
clones.  It accounts for exponentially decaying lifetime thymic
export, a constant rate of cellular proliferation, and variable
cellular death rate that adjusts to present cell counts and
availability of survival resources.  It depicts the generation of the
naive T-cell pool in early life via thymic export, and long-term
maintenance of the population via peripheral turnover after thymic
export has waned.  Values of most of the model's parameters can be
found in previous literature, while the few exceptions are obtained by
fitting some basic results of the model, such as age-related T-cell
loss, to previous experiments. Our analysis serves two important
purposes: to map the thymic machinery, identifying which components do
and do not contribute to age-related cellular loss, and then to
interpret the nuanced role of that cellular loss in immunosenescence.

First, we have found that if thymic export is assumed to decay
exponentially to zero, then all compartments $c_k(t)$ (with $1 \leq k
\leq M$) deplete as $t \to \infty$, independent of essentially any
restrictive assumptions about the homeostatic proliferative mechanism
in the periphery.  Concretely, for any choice of proliferation and
death rates $p(N), \mu(N)$, that satisfy $p(0), \mu(0) > 0$ and the
choice $\gamma(t) = \gamma_0 e^{-at}$ with $\gamma_0, a > 0$, there
exists a sufficiently small $\delta > 0$ guaranteeing $c_k(t) \to 0$
as $t \to \infty$ for all $1 \leq k \leq M$, provided that $\sum
|c_k(1)| \leq \delta$.  Although this result only guarantees that
trajectories $c_k(t)$ started sufficiently close to zero converge to
zero, simulation indicates that the basin of attraction to this ``zero
state" is actually quite large.  In fact, for the typical initial
conditions used throughout this paper, simulation suggests convergence
of all compartments $c_k$ to zero in infinite time.  Although it takes
an extremely long time to deplete all $c_k$ compartments
for $1 \le k \le M$,
the initial phase of this process can still cause
significant loss of T-cell diversity in aging individuals within a
human lifespan. Most importantly, we find that the T-cell
loss driven by
exponentially-diminishing thymic export \textit{alone} is robust
against any assumptions about the homeostatic proliferative mechanism
in the periphery, as this outcome is universal for all functional
forms of $p(N), \mu(N)$; even a particularly strong homeostatic
mechanism (say, one with $p(0) \gg \mu(0)$) cannot rescue a plunging
diversity.  This, in turn, suggests that in searching for treatments
of age-induced loss of diversity, efforts should be directed at the
thymus, in particular to maintaining thymic productivity into advanced
age.

Moreover, we compare the real-time simulations and the
quasisteady-state solutions of the total cell count, as well as the
number of distinct clones, over the course of age-related thymic
output erosion. We find that our simulation results keep lagging
behind the quasi steady state solutions, suggesting that the erosion
time scale of thymic output is faster than the time scale for the
population dynamics to relax towards a steady state. Mathematically,
this result reveals that the evolution of the T-cell population within the
human lifespan is a rather dynamical phenomenon, which may not be
well-described by quasistatic solutions, requiring evaluation of the
fully nonautonomous system.  Biologically, our results indicate that
the loss of T-cell diversity is a delayed response to thymic
involution, and assessment of thymic function
may predict the health of the immune system.

\begin{figure}[h]
    \centering
    \includegraphics[width=\textwidth]{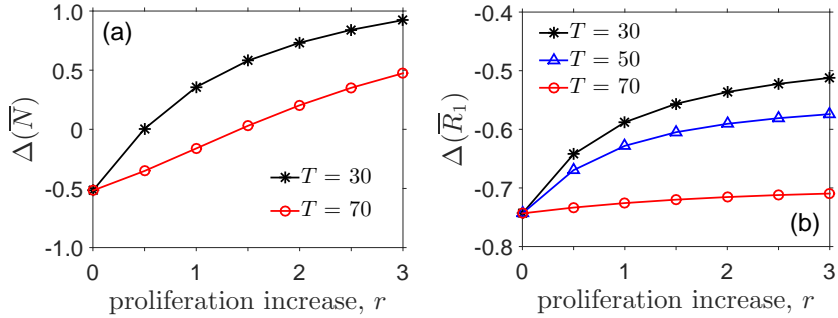}
    \captionsetup{name=Fig.}
    \caption{{\bf Total Cell Count and Richness with Rise in
        Proliferation.}  Simulation of Eq.~\ref{eq:CLONEEQ2} with
      exponentially decaying thymic export, and peripheral homeostasis
      described by time-varying logistic growth.  We use the thymic
      export rate $\gamma(t) = \gamma_0 e^{-at}$, peripheral death
      rate $\mu(N) = \mu_0 > 0$, and peripheral proliferation rate
      $p(N,t) = p(t)(1 - (N/K))$, with $p(t) = p_0 (1 + r H(t-T))$.
      Here, $H(t)$ represents the Heaviside function with jump at
      $t=0$.  The constant $r$ determines the magnitude of the
      increase to the basal proliferation rate, and $T$ represents the
      time at which the jump occurs.  We take the jump to occur at
      varying ages. (a) $\Delta(\bar{N})$ with jump at ages $T=30$ and
      70, for varying $r$.  (Curve corresponding to $T=50$ is omitted
      due to close similarity to $T=30$ curve.)  Raising the basal
      proliferation rate diminishes cellular loss in advanced age,
      with sufficiently high values of $r$ producing a lifetime
      increase in total cell counts.  The positive steady state
      solution of the autonomous total cell ODE, ${\rm d}N/{\rm d}t =
      \gamma_0 + p_0(1 - N/K) - \mu_0 N$, is given by $N^* = (K/2) (1
      - \mu_0/p_0 + \sqrt{(1 - \mu_0/p_0)^2 + 4 \gamma_0 / K p_0})$,
      and can be seen to satisfy $\partial N^*/ \partial p_0 > 0$ if
      $\gamma_0 < K \mu_0$, suggesting that increases to the basal
      proliferation rate are likely to increase the total cell count.
      (b) $\Delta(\bar{R}_1)$ with $T=30, 50,$ and $70$, for varying
      $r$.  Increases to the basal proliferation rate do mitigate
      diversity loss, but the effect is minor and potentially
      insignificant.  Increases to the basal proliferation rate
      increase $c_{k+1}$ due to a decrease in $c_k$, preserving
      additional diversity, but the lifetime diversity loss is still
      observed, even when proliferation rates are high enough to
      generate a lifetime increase to the total cell count.  Fixed
      parameter values: $\gamma_0 = 1.8 \times 10^{10}$, $a = 0.044$,
      $p_0 = 0.18$, $\mu_0 = 0.17$, $K_0 = 3 \times 10^{11}$, $\Omega
      = 10^{16}$. Initial values: $c_0(1) = \Omega - 10^{11}$, $c_1(1)
      = 10^{11}$ $c_k(1) = 0$ for $k \geq 1$. Eq.~\ref{eq:CLONEEQ2} is
      truncated at $k=200$.}
    \label{fig:2XPROLIF}
\end{figure}

Although peripheral division cannot salvage the T-cell population on a
long time scale, higher basal proliferation rates may at least delay
the erosion of the T-cell compartment, sustaining acceptable
effectiveness of the immune system within the human
lifespan~(\citealt{NAYLOR2005}).  We assumed a constant lifetime rate
of cellular proliferation, but alternative research suggests that
proliferation rates may increase with age~(\citealt{NAYLOR2005}).  In
light of this finding, we briefly inspect the effect of increased
proliferation rates at advanced ages on cellular and clonal loss by
modifying $p(N)$ and $\mu(N)$ in
Eq.~\ref{eq:CLONEEQ2}. For simplicity, we take the
death rate to be constant ($\mu(N) = \mu_0 > 0$), and adopt a logistic
growth rate, $p(N,t) = p(t)(1-N/K)$, where a discrete increase in the
proliferation rate is incorporated in $p(t) = p_0(1 + r H(t-T))$, with
$p_0 > 0$ the early-life basal cellular proliferation rate, and $H(t)$
the Heaviside function, with $T$ the age at which the rate increases.
The constant $r$ specifies the increase to the proliferation rate.
(Full simulation details are given in the caption of
Fig.~\ref{fig:2XPROLIF}.)  By varying $r$, simulation under these
alternate hypotheses indicates that increased basal proliferation
rates do lead to notably higher total cell counts
(Fig.~\ref{fig:2XPROLIF}(a)), but have little effect on diversity
(Fig.~\ref{fig:2XPROLIF}(b)).  These results further affirm that
expansion of peripheral proliferation is unlikely to rescue the
eroding naive T-cell diversity, despite the increased cell count.
If diversity loss is the main cause of immunosenescence (still a
debatable topic in the medical community), peripheral proliferation
may not be the sensible target of treatments.

The increased $N(70)$ and nearly unchanged $R_{1}(70)$ in
Fig.~\ref{fig:2XPROLIF} imply that the decline of T-cell diversity at
old age may appear more dramatic if the diversity is measured in terms
of frequency of distinct TCR sequences among the cycling cells, which
corroborates the explanation that an increase of proliferation rate at
old age leads to a sharp decrease of T-cell
diversity~(\citealt{NAYLOR2005}). Previous models have shown that even
sharper decline of T-cell diversity can be induced by fitness
selection, where certain clonotypes increase their fitness at old age
possibly due to higher avidity to
self-antigens~(\citealt{JOHNSON2012,JOHNSON2014,GORONZY2015b}).

Although the boosts to the total cell count through artificial
expansion of the proliferative mechanism are unable to replenish the
declining TCR diversity in the naive T-cell pool, it is possible that
the impact is less severe than the decaying richness would have
indicated, considering that most of the extinct clones are originally
small clones, which may be much less effective than larger clones.  In
this regard, the viability of treating immunosenescence by expanding
peripheral proliferation depends on the elucidation of the T-cell
pool's \textit{effectiveness clone size}--that is, the size a clone
must have attained to effectively guarantee activation of the clone
when its cognate antigen infiltrates the organism.  The effectiveness
clone size is intrinsically linked to true functional TCR diversity;
if we can identify a threshold integer $q^*$, such that clones of size
at least $q^*$ are reliably activated in the presence of their cognate
antigen(s), but that smaller clones are not, then $R_{q^*}(t)$ is
naturally the most useful measure of diversity, because it accounts
for precisely those clones actively participating in the adaptive
immune mechanism.  The larger the ``correct" choice of $q^*$ is, the
more effective treatments to boost cellular proliferation in the
periphery will be.  Our model directly yields the number of clones of
a particular size, making it straightforward to include or exclude
clones below a certain cell count, should such a threshold exist and
be identified.

The effective clone size is also significant to the question of
whether diversity loss is the driving factor in immunosenescence.
Using the parameter values that we found in the literature, $R_q(t)$
decreases for $q \leq 4$ from youth to advanced age, stays nearly
constant for $q = 5, 6$, and increases for $q \geq 7$.  The extinction
of small clones allows the surviving clones to expand in size, leading
the richness of large clones to increase at old ages.  If the minimal
size for a T-cell clone to effectively respond to antigens is large,
the diversity of such ``effective'' clones may actually increase with
age, strengthening the immune response.  Therefore, either the minimal
clone size required for effective immune response is low, or the
weakened immune response at old ages is caused primarily by other
mechanisms.  For example, functional deficiencies acquired by naive
T-cells in aging are one possible alternative cause of the weakened
immune response.  Such functional deficiencies have been studied
heavily in mouse models, but research in humans is still
lacking~(\citealt{APPAY2014}).  Diminished naive T-cell effector
responsiveness and proliferative capacity have been observed in aged
mice~(\citealt{MORO-GARCIA2013}).  It is possible that similar changes
occur in humans.  Conversely, experiments on mice have directly shown
that loss of TCR diversity does have an actively detrimental effect on
immune responsiveness~(\citealt{YAGGER2008}), supporting the notion
that loss of TCR diversity as a significant contributor to
immunosenescence.

Our model illustrates the feasibility of several different scenarios,
in which loss of diversity contributes to immunosenescence on
drastically different levels.  There is clearly a strong need to
investigate the effects of both age-related structural diversity loss
and T-cell functionality loss in human subjects \textit{in vivo}, to
better understand the causes of immunosenescence.  Moreover, our model
indicates that the effectiveness clone size and crossreactivity
\textit{in vivo} are valuable pieces of missing information, the
elucidation of which would allow for the identification of effective
options to treat immunosenescence.

\section{Summary and Conclusions}
\label{sec:4}
We have simulated the time evolution of the functions $c_k(t)$, which
represent the number of naive T-cell clones of size $k$ present in a
human's immune compartment at time $t$.  We determined that under
essentially any realistic assumptions about homeostatic proliferation
and death, all clones deplete in infinite time if thymic export is
assumed to decay exponentially.  This implicates thymic export as a
fundamental cause of age-associated diversity loss.  We simulated our
model under the assumption that a carrying capacity is regulated by
homeostatic proliferation and death through $N$-dependent rates.  We
found that the manipulation of homeostatic proliferation and death
rates, which may notably raise the carrying capacity and thus the
total cell count, was unable to save falling diversity as an
individual ages.  It affirms the vital role of thymic output in
age-related diversity loss, and indicates that boosting the
proliferation rate is unlikely an effective solution.  However, if
only clones of large size are sufficiently effective in the immune
response, boosting proliferation rates might raise average clone sizes
and help to mitigate the effects of lost diversity.  We simulated
``threshold richness diversity", $R_q(t)$, which counts the total
number of clones of size $q$ or larger.  We found that by increasing
$q$, the trajectory of $R_q(t)$ changes from decreasing to increasing
over a human lifetime. From this trend, we concluded that if only
large clones are effective, the effective richness would actually
increase with age, suggesting that it is important to identify the
minimal effective clone size in order to determine whether the loss of
TCR diversity is the primary driving mechanism of the immune
dysfunction seen in advanced age. Lastly, we derived a one-to-one
mapping between the full-sample diversity $c_k^N$ of $N$ cells and the
expected measurement of diversity $\mathbb{E}[c_k^Y]$ in samples of
$Y$ cells. We found that the probability of detecting small clones
shrank significantly with small sample sizes, which could potentially
skew small-sample statistics.  In particular, we show that small
samples tend to underestimate the age-related loss of T-cell richness
diversity.  Our formulation provides a rigorous method for accurately
inferring the statistical distribution of clonal sizes from
small-sample measurements.

\section*{Acknowledgements}
This work was supported by the NIH (SL, R56HL126544), the NSF (TC and
SL, DMS-1516675 and DMS-1814364) and the Army Research Office (YLC,
W1911NF14-1-0472).

\appendix

\section{Implementation of Numerical Truncation}
\label{sec:APP1}

The most straightforward way to truncate Eq.~\ref{eq:CLONEEQ2} at $k =
M$ is to neglect the exchange terms between $c_{M}$ and $c_{M+1}$,
assuming a negligible contribution for $k > M$ and essentially
imposing a ``no-flux" boundary condition.  This leads to the following
equation for the boundary term $c_{M}(t)$:

\begin{align}
\frac{{\rm d}c_M(t)}{{\rm d}t} &= \frac{\gamma(t)}{\Omega}c_{M-1} +
p(M-1)c_{M-1} - \mu(N) M c_M. \label{eq:TRUNCATED3}
\end{align}

\noindent
This formulation, however, introduces a truncation error in
Eq.~\ref{eq:EQNN} if we define $N = \sum_{k=1}^M k c_k$.  The
neglected terms leave a small loss of total cell count in ${\rm d} N /
{\rm d} t$.  An alternative implementation of the truncation is adding
these small loss terms to the boundary equation:

\begin{align}
\frac{{\rm d}c_M(t)}{{\rm d}t} &= \frac{\gamma(t)}{\Omega}
\left( c_{M-1} + \frac{c_M}{M} \right)
+ p(M-1)c_{M-1} + p c_M - \mu(N) M c_M, \label{eq:TRUNCATED4}
\end{align}

\noindent
thus preserving the total cell count $N$. However, for
Eq.~\ref{eq:TRUNCATED4} the truncation error shows up in the total
number of clonal types $\Omega = \sum_{k=0}^M c_k$, as the terms added
to Eq.~\ref{eq:TRUNCATED4} to preserve $N$ artificially introduce new
clonal types into the model. In contrast, $\Omega$ is preserved with
the implementation of Eq.~\ref{eq:TRUNCATED3}.  If $M \to \infty$, the
truncation errors for both implementations go to zero at $\sim 1 / M$,
and the two implementations become equivalent.  Assuming sufficiently
large $M$, the truncation errors can be negligible in the context of
$\gamma (t) > 0$, or have minimal cumulative effects within a limited
duration, such as a human lifetime, on which our investigations in
this paper have primarily focused.

In this paper, we adopt, for simplicity, Eq.~\ref{eq:TRUNCATED3} to
numerically truncate Eq.~\ref{eq:CLONEEQ2}. Note that this choice may
seem ``natural'' if one regards $M$ as the carrying capacity, making
it reasonable for $c_M$ to have zero proliferation rate. However, the
full mechanisms associated with the carrying capacity are far more
sophisticated than simply eliminating the proliferation of $c_M$. Not
only should the proliferation rate of $c_M$ go to zero, the
proliferation rate of the other $c_k$ should also have a $k$
dependence. The $k$ dependence may be weak for small $k$, but as $k
\to M$, the proliferation rate should attenuate significantly. The
probability that $c_{k \to M}$ will proliferate should be very small,
as it is highly likely that there exist other smaller clones to push
the total cell count up to the carrying capacity, prohibiting further
proliferation. The $k$-dependent proliferation rate will yield a
natural truncation threshold at the carrying capacity. However, such a
sophisticated $k$-dependence of the proliferation rate is beyond the
scope of this paper.  Our assumption here is simply that the
truncation errors introduced by Eq.~\ref{eq:TRUNCATED3} are
numerically negligible and not biologically significant.

\section{Steady States of the Autonomous Equations}
\label{sec:APP2}

If we fix $\gamma(t) = \gamma_0$,
Eqs.~\ref{eq:EQNN},~\ref{eq:CLONEEQ2}, and~\ref{eq:TRUNCATED3} become
autonomous and admit the following steady state solution,

\begin{align}
c_1^{\rm ss} &= \gamma_0 \left[\frac{\gamma_0}{\Omega} \sum_{i = 1}^M \frac{1}{i! \mu(N_{\rm ss})^{i-1}}
\left(\prod_{j=1}^{i-1} \left[\frac{\gamma_0}{\Omega} + j p\right] \right) + \mu(N_{\rm ss})\right]^{-1},\label{eq:SS1}\\
c_k^{\rm ss} &= \frac{c_1^{\rm ss}}{k! \mu(N_{\rm ss})^{k-1}} \left(\prod_{n=1}^{k-1}\left[\frac{\gamma_0}{\Omega} + n p\right]\right),\label{eq:SS2}
\end{align}

\noindent where $N_{\rm ss}$ is the total population at steady state,
given by the unique positive root of the cubic,

\begin{align}
\label{eq:CUBIC}
c(N; \gamma_0) = \left(p_0 - (\mu_0 + \mu_1)\right) N^3 + \gamma_0 N^2 + (p_0 - \mu_0) K^2 N + \gamma_0 K^2.
\end{align}

\noindent When $\gamma_0 = 0$, $c(N; 0)$ has three real roots, $N =
0$, $\pm \sqrt{((p - \mu_0) K^2)/(\mu_0 + \mu_1 - p)}$.  The positive
steady state solution, which we denote by $N_{\rm ss}(0)$, is stable,
and the zero solution unstable, under the parameter restrictions
described in Section~\ref{sec:2.1}.  We now demonstrate that even
though Eqs.~\ref{eq:SS1},~\ref{eq:SS2} indicate that each $c_k^{\rm
  ss} \to 0$ as $\gamma_0 \to 0$, the quantity $\lim_{M \to \infty}
\sum_{k=1}^M k c_k^{\rm ss}$ converges to a positive value
qualitatively consistent with $N_{\rm ss}(0)$ as $\gamma_0 \to 0$.

\vskip6pt

\noindent \textbf{Proposition B:} The steady state solutions $c_k^{\rm ss}$,
as given in Eqs.~\ref{eq:SS1},~\ref{eq:SS2}, satisfy,

\vskip4pt

\begin{align*}
\lim_{\gamma_0 \to 0} \lim_{M \to \infty} \sum_{k=1}^M k c_k^{\rm ss} > 0
\end{align*}

\vskip8pt

\textit{Proof.}  We seek to derive upper and lower bounds,
$U(\gamma_0)$, $L(\gamma_0)$, which satisfy,

\begin{align*}
L(\gamma_0) \leq \lim_{M \to \infty} \sum_{k=1}^M k c_k^{\rm ss} \leq U(\gamma_0),
\end{align*}

\noindent for $\gamma_0 > 0$, and $\lim_{\gamma_0 \to 0} U(\gamma_0)
\geq \lim_{\gamma_0 \to 0} L(\gamma_0) > 0$.  We first establish two
small results, which will be used later on:

\vskip20pt

\noindent \textbf{Proposition B1:} For $\mu = \mu(N_{\rm
  ss}(\gamma_0))$, $\lim_{\gamma_0 \to 0} \frac{{\rm d}\mu}{{\rm
    d}\gamma_0} > 0$.

\vskip4pt

\textit{Proof.}  Recalling that $\mu = \mu(N_{\rm ss}(\gamma_0)) =
\mu_0 + \mu_1 (N_{\rm ss}(\gamma_0)^2/(N_{\rm ss}(\gamma_0)^2 +
K^2))$, we have:

\vskip6pt

\begin{align*}
\frac{{\rm d}\mu}{{\rm d}\gamma_0} &= \frac{{\rm d}\mu}{{\rm d} N_{\rm ss}} \frac{{\rm d}N_{\rm ss}}{{\rm d}\gamma_0}\\
&= \frac{2 \mu_1 K^2 N_{\rm ss}}{(N_{\rm ss}^2 + K^2)^2} \left[ \frac{- (N_{\rm ss}^2 + K^2)}{3 (p_0 - (\mu_0 + \mu_1))N_{\rm ss}^2 + 2 \gamma_0 N_{\rm ss} + (p_0 - \mu_0)K^2}\right]\\
&= \frac{-2 \mu_1 K^2 N_{\rm ss}}{(N_{\rm ss}^2 + K^2)\left[3 (p_0 - (\mu_0 + \mu_1))N_{\rm ss}^2 + 2 \gamma_0 N_{\rm ss} + (p_0 - \mu_0)K^2\right]}
\end{align*}

\vskip10pt

\noindent where we computed the derivative $\frac{{\rm d}N_{\rm
    ss}}{{\rm d}\gamma_0}$ implicitly from the expression $c(N_{\rm
  ss}(\gamma_0); \gamma_0) = 0$.  From the explicit form $N_{\rm
  ss}(0) = \sqrt{(p_0 - \mu_0) K^2 / ((\mu_0 + \mu_1) - p_0)}$, we
have:

\vskip6pt

\begin{align*}
\lim_{\gamma_0 \longrightarrow 0} \frac{{\rm d}\mu}{{\rm d}\gamma_0} &= \frac{-2 \mu_1 K^2 N_{SS}(0)}{(N_{SS}(0)^2 + K^2)\left[3 (p_0 - (\mu_0 + \mu_1))N_{SS}(0)^2 + (p_0 - \mu_0)K^2\right]}\\
&= \frac{-2 \mu_1 K^2 N_{SS}(0)}{(N_{SS}(0)^2 + K^2)\left[-2(p_0 - \mu_0) K^2\right]}\\
&> 0
\end{align*}

\vskip10pt

\noindent \textbf{Proposition B2:} For $f(p/\mu(N_{\rm ss}(\gamma_0));
\gamma_0) = \frac{\gamma_0}{p \Omega} \left( 1 - \frac{p}{\mu(N_{\rm
    ss}(\gamma_0))}\right)^{\frac{-\gamma_0}{p \Omega} - 1}$,
$\lim_{\gamma_0 \to 0} f(p/\mu(N_{\rm ss}(\gamma_0)); \gamma_0) > 0$.

\vskip4pt

\textit{Proof.}  We write the function $f(p/\mu(N_{\rm ss}(\gamma_0));
\gamma_0)$ as a product of two functions as follows:

\vskip6pt

\begin{align*}
f(p/\mu(N_{\rm ss}(\gamma_0)); \gamma_0) &= \frac{\gamma_0}{p \Omega} \left( 1 - \frac{p}{\mu(N_{\rm ss}(\gamma_0))}\right)^{\frac{-\gamma_0}{p \Omega} - 1}\\
&= \left(1 - \frac{p}{\mu(N_{\rm ss}(\gamma_0))}\right)^{\frac{-\gamma_0}{p \Omega}} \cdot \frac{\gamma_0}{p \Omega} \left(1 - \frac{p}{\mu(N_{\rm ss}(\gamma_0))}\right)^{-1}\\
&= A(\gamma_0) \cdot B(\gamma_0)
\end{align*}

\vskip6pt

\noindent We define $A_0 = \lim_{\gamma_0 \to 0} A(\gamma_0)$ and $B_0
= \lim_{\gamma_0 \to 0} B(\gamma_0)$, and compute $A_0$ and $B_0$:

\vskip6pt

\begin{align*}
\ln(A_0) &= \lim_{\gamma_0 \longrightarrow 0} \frac{-\gamma_0}{p \Omega} \ln\left(1 - \frac{p}{\mu(N_{\rm ss}(\gamma_0))}\right)\\
&= \frac{-1}{p\Omega} \lim_{\gamma_0 \longrightarrow 0} \frac{\ln\left(1 - \frac{p}{\mu(N_{\rm ss}(\gamma_0))}\right)}{\gamma_0^{-1}}\\
&= \frac{-1}{p\Omega} \lim_{\gamma_0 \longrightarrow 0} \frac{\left(1 - \frac{p}{\mu(N_{\rm ss}(\gamma_0))}\right)^{-1} \frac{\rm d}{{\rm d}\gamma_0}\left(- \frac{p}{\mu(N_{\rm ss}(\gamma_0))}\right)}{-\gamma_0^{-2}}\\
&= \frac{1}{p\Omega} \lim_{\gamma_0 \longrightarrow 0}  \gamma_0^2 \left[ 1 - \frac{p}{\mu(N_{\rm ss}(\gamma_0))} \right]^{-1} \left[ p \mu(N_{\rm ss}(\gamma_0))^{-2} \frac{{\rm d}\mu}{{\rm d}\gamma_0}\right]\\
&= \frac{1}{p\Omega} \lim_{\gamma_0 \longrightarrow 0}  \left[\frac{\gamma_0^2 p \frac{{\rm d}\mu}{{\rm d}\gamma_0}}{\mu(N_{\rm ss}(\gamma_0))^2 - p \mu(N_{\rm ss}(\gamma_0))}\right]\\
&= \frac{1}{\Omega} \lim_{\gamma_0 \longrightarrow 0} \left[ \frac{2 \gamma_0 \frac{{\rm d}\mu}{{\rm d}\gamma_0} + \gamma_0^2 \frac{{\rm d}^2 \mu}{{\rm d}\gamma_0^2}}{(2\mu - p)\frac{{\rm d}\mu}{{\rm d}\gamma_0}}\right]\\
&= \frac{1}{\Omega} \left[ \frac{2 \gamma_0 \lim_{\gamma_0 \to 0} \frac{{\rm d}\mu}{{\rm d}\gamma_0} + \gamma_0^2 \lim_{\gamma_0 \to 0} \frac{{\rm d}^2 \mu}{{\rm d}\gamma_0^2}}{p \lim_{\gamma_0 \to 0} \frac{{\rm d}\mu}{{\rm d}\gamma_0}}\right],
\end{align*}

\vskip6pt

\noindent where we used that $\mu(N_{\rm ss}(\gamma_0)) \to p$ as
$\gamma_0 \to 0$.  From Proposition B1, $\lim_{\gamma_0
  \longrightarrow 0} \frac{{\rm d}\mu}{{\rm d}\gamma_0} > 0$, and a
similar computation shows that $\lim_{\gamma_0 \to 0} \frac{{\rm d}^2
  \mu}{{\rm d} \gamma_0^2} \in \mathbf{R}$.  Thus, $\ln(A_0) \in
\mathbf{R}$, and $A_0 > 0$.  Now,

\vskip6pt

\begin{align*}
B_0 &= \lim_{\gamma_0 \to 0} \frac{\gamma_0}{p \Omega} \left(1 - \frac{p}{\mu(N_{\rm ss}(\gamma_0))}\right)^{-1}\\
&= \lim_{\gamma_0 \to 0} \frac{(\gamma_0/p \Omega)}{\left(1 - \frac{p}{\mu(N_{\rm ss}(\gamma_0))}\right)}\\
&= \lim_{\gamma_0 \to 0} \frac{(1/p \Omega)}{p \mu(N_{\rm ss}(\gamma_0))^{-2} \frac{{\rm d}\mu}{{\rm d}\gamma_0}}\\
&= \lim_{\gamma_0 \to 0} \frac{\mu(N_{\rm ss}(\gamma_0))^2 }{p^2 \Omega \frac{{\rm d}\mu}{{\rm d}\gamma_0}}\\
&> 0.
\end{align*}

\vskip6pt

\noindent Thus, $\lim_{\gamma_0 \to 0} \frac{\gamma_0}{p \Omega}
\left( 1 - \frac{p}{\mu(N_{\rm ss})}\right)^{\frac{-\gamma_0}{p
    \Omega} - 1} = A_0 B_0 > 0$.

\vskip6pt

\noindent We now resume the proof of Proposition B.  We first derive
upper and lower bounds on the term $c_1^{\rm ss}$, to simplify
calculations.  From the nonnegativity of the parameters and
coefficient functions, and the form in Eq.~\ref{eq:SS1}, $c_1^{\rm ss}
\leq \gamma_0/\mu_0$, independent of $M$.  To derive an
$M$-independent lower bound on $c_1^{\rm ss}$, we observe that the sum
in the denominator of Eq.~\ref{eq:SS1} satisfies,

\vskip6pt

\begin{align*}
\frac{\gamma_0}{\Omega} \sum_{i=1}^M \frac{1}{i! \mu(N_{\rm ss}(\gamma_0))^{i-1}} 
\left(\prod_{j=1}^{i-1} \left[ \frac{\gamma_0}{\Omega} + jp\right]\right) &\leq 
\sum_{i=1}^M \frac{1}{(i-1)! \mu(N_{\rm ss}(\gamma_0))^{i-1}} \left(\prod_{j=0}^{i-1} 
\left[ \frac{\gamma_0}{\Omega} + jp\right]\right)\\
&= p \sum_{i=1}^M \frac{1}{(i-1)!} \left(\prod_{j=0}^{i-1} \left[ \frac{\gamma_0}{p \Omega} + j\right]\right) \left(\frac{p}{\mu(N_{\rm ss}(\gamma_0))}\right)^{i-1}
\end{align*}

\vskip6pt

\noindent and that the sum on the right above is the $M$-th Taylor
polynomial, $S_{M, \gamma_0}$, for the function $f(x; \gamma_0) =
\frac{\gamma_0}{p \Omega} \left( 1- x\right)^{\frac{-\gamma_0}{p
    \Omega} - 1}$ expanded around $x = 0$ and evaluated at $x =
\frac{p}{\mu(N_{\rm ss}(\gamma_0))}$.  The function $f(x; \gamma_0)$
is analytic in $x$ away from $x = 1$, and in particular, the $S_{M,
  \gamma_0}$ increase monotonically to $f(p/\mu(N_{\rm ss}(\gamma_0));
\gamma_0)$.  It follows that,

\vskip6pt

\begin{equation*}
\frac{1}{p} \sum_{i=1}^M \frac{1}{i! \mu(N_{\rm ss}(\gamma_0))^{i-1}} \left(\prod_{j=0}^{i-1} \left[ \frac{\gamma_0}{\Omega} + jp\right]\right) \leq S_{M,\gamma_0} \leq f \left(\frac{p}{\mu(N_{\rm ss}(\gamma_0))}; \gamma_0\right) := f_{\gamma_0}
\end{equation*}

\vskip6pt

\noindent and thus that $c_1^{\rm ss} \geq \gamma_0/(p f_{\gamma_0} +
\mu_0 + \mu_1)$.  After using the $c_1^{\rm ss}$ bounds in the
expression for $c_k^{\rm ss}$, we have:

\vskip6pt

\begin{align*}
&\frac{p \Omega}{p f_{\gamma_0} + \mu_0 + \mu_1} S_{M,\gamma_0} \leq \sum_{k=1}^M k c_k^{\rm ss} \leq \frac{p \Omega}{\mu_0} S_{M,\gamma_0}\\
\longrightarrow &\lim_{M \to \infty} \frac{p \Omega}{p f_{\gamma_0} + \mu_0 + \mu_1} S_{M,\gamma_0} \leq \lim_{M \to \infty} \sum_{k=1}^M k c_k^{\rm ss} \leq \lim_{M \to \infty} \frac{p \Omega}{\mu_0} S_{M,\gamma_0}\\
\longrightarrow &\frac{p \Omega}{p f_{\gamma_0} + \mu_0 + \mu_1} f_{\gamma_0} \leq \lim_{M \to \infty} \sum_{k=1}^M k c_k^{\rm ss} \leq \frac{p \Omega}{\mu_0} f_{\gamma_0}
\end{align*}

\vskip6pt

\noindent Now we let $L(\gamma_0) = \frac{p \Omega}{p f_{\gamma_0} +
  \mu_0 + \mu_1} f_{\gamma_0}$ and $U(\gamma_0) = \frac{p
  \Omega}{\mu_0} f_{\gamma_0}$.  From Proposition B2, $\lim_{\gamma_0
  \to 0} f_{\gamma_0} > 0$, so $\lim_{\gamma_0 \to 0} L(\gamma_0),
\lim_{\gamma_0 \to 0} U(\gamma_0) > 0$, and Proposition B follows.

\section{Convergence and Stability of $c_k$ when $\gamma (t) \to 0$}
\label{sec:APP3}

\noindent In this section we will prove that solutions $c_{k}$ to our
ODE system initialized sufficiently close to
$\mathbf{\overrightarrow{0}}$ converge to
$\mathbf{\overrightarrow{0}}$ as $t \to \infty$.  Denote by (P) the
``perturbed" ODE system given by
Eqs.~\ref{eq:CLONEEQ2},~\ref{eq:TRUNCATED3}, with $\gamma(t) =
\gamma_0 e^{-at}$, and by (U) the ``unperturbed" ODE system resulting
from the alternate choice $\gamma(t) \equiv 0$. For the sake of
generality, we omit previous assumptions about the form of the
functions $p(N), \mu(N)$, except that $p(0), \mu(0) > 0$.
Additionally, in this section, we regard the term $N$ that appears in
the ODEs as $\sum_{k \geq 1} k c_k$ instead of its own variable, and
thus do not explicitly include Eq.~\ref{eq:EQNN} in our analysis as in
Appendix~\ref{sec:APP2}. Note that the residual $N - \sum_{k \geq 1} k
c_k \to 0$ as $M \to \infty$.  We begin by noting that the unperturbed
system (U) has steady-state $c_k^U(t) \equiv 0$ for $k \geq 1$.  To
analyze the stability of this steady state, we consider the
linearization of (U) around this steady state, which is represented by
the $M \times M$ matrix we call $\mathbf{L_U}$ ($\mathbf{L_U} =
(l_{ij})_{1 \leq i,j \leq M}$).  The components $l_{ij}$ of
$\mathbf{L_U}$ are given explicitly by:

\begin{align} \label{eq:LU}
\begin{displaystyle}
  l_{ij} = \left.
  \begin{cases}
    -j(p(0)+\mu(0)), & \text{if } i = j \leq M-1 \\
    -M \mu(0), & \text{if } i = j = M \\
    j \mu(0), & \text{if } i = j - 1; \text{  } 2 \leq j \leq M\\
    j p(0), & \text{if  } i = j + 1; \text{  } 1 \leq j \leq M-1\\
    0, & \text{otherwise  }
  \end{cases}
  \right\}
\end{displaystyle}
\end{align}

\noindent Although the matrix is tridiagonal, it is high-dimensional,
and thus its eigenvalues cannot be computed analytically.  However, we
may nevertheless demonstrate that all eigenvalues possess strictly
negative real part, indicating that the zero solution is
asymptotically stable.  To do this, we use Gershgorin's circle theorem
to show that if there exists an eigenvalue $\lambda \in \mathbf{C}$
satisfying $\Re(\lambda) \geq 0$, then $\lambda = 0$.  We then verify
that $\lambda = 0$ is never an eigenvalue of $\mathbf{L_U}$, by
directly demonstrating that $\mathbf{L_U}$ has linearly independent
rows.

\vskip12pt

\noindent \textbf{Proposition C:} \textit{All eigenvalues $\lambda \in
  \mathbf{C}$ of the matrix $\mathbf{L_U}$ satisfy $\Re{\lambda} < 0$, so that
  the zero-solution of (U) is asymptotically stable.}

\vskip12pt

We first apply Gershgorin's circle theorem to the columns of the
matrix $\mathbf{L_U}$ to conclude that all eigenvalues $\lambda \in
\mathbf{C}$ of the truncated system (finite $M$) are contained within
the following union of disks:

\vskip8pt

\begin{equation}
\left(\bigcup_{i=1}^{M-1} \{\lambda \in \mathbf{C} : |\lambda + i (p(0) + \mu(0))| \leq i (p(0)+\mu(0))\}\right) \bigcup \text{  } \{ \lambda \in \mathbf{C}: | \lambda + M \mu(0)| \leq M \mu(0)\},
\label{eq:GERSHGORIN}
\end{equation}

\vskip8pt

\noindent where we have used the fact that $\{\lambda \in \mathbf{C}:
|\lambda + D| \leq D\} \subset \{\lambda \in \mathbf{C}: |\lambda + (D
+ \epsilon)| \leq D + \epsilon\}$ for $D, \epsilon > 0$.  Given the
assumption that $p(0), \mu(0) > 0$, each of these disks is tangent to
the line $\Re \lambda = 0$ at $\lambda = 0$, and otherwise lies
entirely in the half plane $\Re \lambda < 0$.  Thus, $\mathbf{L_U}$
can only possess an eigenvalue $\lambda$ satisfying $\Re \lambda = 0$
if $\lambda = 0$ is itself an eigenvalue.  We next verify that
$\lambda = 0$ is never an eigenvalue of $\mathbf{L_U}$ directly, by
establishing the linear independence of the rows of $\mathbf{L_U}$.

\vskip 8pt

Let us assume that there exist scalars $a_1, a_2, \ldots, a_M$, such
that $\sum_{j = 1}^M a_j \left( l_{ij} - 0 \right) = 0$ for all $1 \le
i \le M$.  Hence a normalized vector $\mathbf{a} = \left( a_1, a_2,
\ldots, a_M \right)$ represents the eigenvector of the zero
eigenvalue.  For $i = 1$, we find that $2 a_2 \mu(0) - a_1 (p(0) +
\mu(0)) = 0$, so that $a_2 = 2^{-1} \mu(0)^{-1} (p(0) + \mu(0)) a_1$.
By moving on to larger $i$, we can recursively derive $a_i = \Theta_i
a_1$ for all $2 \le i \le M$ with a proportional constant coefficient
$\Theta_i$.  Moreover, $\sum_{i = 1}^M \sum_{j = 1}^M a_j l_{ij} = -
a_1 \mu(0) = 0$, leading to $a_1 = 0$ given that $\mu(0) > 0$. If $a_1
= 0$, $\mathbf{a} \equiv 0$, and a non-zero eigenvector does not
exist, implying that zero is not among the eigenvalues of the $M
\times M$ matrix $\mathbf{L_U}$.  We thus conclude that all
eigenvalues $\lambda$ of the matrix $\mathbf{L_U}$ satisfy
$\Re(\lambda) < 0$, and the zero-solution of (U) is asymptotically
stable for Eq.~\ref{eq:CLONEEQ2} truncated using
Eq.~\ref{eq:TRUNCATED3} at an arbitrarily large $M$.  Note that the
proof in Eq.~\ref{eq:GERSHGORIN} does not hold if we use the
alternative truncation formula Eq.~\ref{eq:TRUNCATED4}.  By forcing
all cells to remain below the truncation threshold $M$, it is not
possible for all $c_k$ to go to zero with a finite $M$.  For the
alternative truncation, the stable steady state solution is $c_k = 2
N_{\rm ss} / (M (M+1))$, which nevertheless goes to zero as $M \to
\infty$.

We next proceed to demonstrate that the uniform asymptotic stability
of the zero-solution ($c_k^U(t) \equiv 0$ for $k \geq 1$) of the
unperturbed system (U) confers a similar notion of ``stability" on the
perturbed system (P).  In particular, the uniform asymptotic stability
of the system (U), in conjunction with the exponential decay of the
function $\gamma(t)$, implies that solutions of the perturbed system
(P) also converge to zero in magnitude, in a sense to be made more
precise later on. Here let us simplify our notation by writing (U) as
$\mathrm{d} \mathbf{c} / \mathrm{d} t = \mathbf{f} \left( \mathbf{c}
\right)$, where $\mathbf{c} \equiv \left( c_1, c_2, \ldots, c_M
\right)$. The autonomous term $\mathbf{f} \left( \mathbf{c} \right)$
consists of cell proliferation and death. Correspondingly we express
(P) as $\mathrm{d} \mathbf{c} / \mathrm{d} t = \mathbf{f} \left(
\mathbf{c} \right) + \mathbf{g} \left( t, \mathbf{c} \right)$, where
the nonautonomous term $\mathbf{g} \left( t, \mathbf{c} \right)$
describes thymic export that depends explicitly on the argument
$t$. We appeal to results of Strauss and Yorke
in~\citeyearpar{STRAUSS1967}, in particular their Theorem 4.6, which
we may invoke to prove that the solution of the perturbed system
$\mathbf{c}^P (t) \to 0$ if the unperturbed and perturbed systems (U)
and (P) satisfy the following conditions:

\vskip8pt

\begin{enumerate}
\item The zero solution ($\mathbf{c}^U(t) \equiv 0$)
  of the unperturbed system (U) is uniformly asymptotically stable.
  \item The autonomous term $\mathbf{f}(\mathbf{c})$ is $C^1$.
  \item There exists $r > 0$ such that if $|\mathbf{c}| \leq r$, then
    $|\mathbf{g}(t,\mathbf{c})| \leq \eta(t)$ for all $t \geq 0$ where $G(t) :=
    \int_{t}^{t+1} \eta(s) ds \to 0$ as $t \to
    \infty$.  (Here, we use the norm $|\mathbf{c}| = \sum_{i=1}^M
    |c_i|$.)
\end{enumerate}

\vskip8pt

\noindent We now verify Conditions 1--3 above.  Condition 1 follows
immediately from the previous discussion, and the fact that for an
autonomous system, asymptotic stability and uniform asymptotic
stability are equivalent. Condition 2 is trivial.  To
verify Condition 3, we must construct a suitable function $\eta(t)$,
using the definition of the function $g(t,\mathbf{c})$:

\vskip8pt

\begin{align}
|\mathbf{g}(t,\mathbf{c})| &= \left|\frac{\gamma_0 e^{-at}}{\Omega}
\left(\Omega - \sum_{j = 1}^M c_j - c_1\right)\right| + \sum_{j =
  2}^{M-2} \left|\frac{\gamma_0 e^{-at}}{\Omega} \left(c_j -
c_{j+1}\right)\right| + \left|\frac{\gamma_0 e^{-at}}{\Omega}
c_{M-1}\right|\\ &\leq \frac{\gamma_0 e^{-at}}{\Omega} \left(|\Omega|
+ \left(\sum_{i = 1}^M |c_i|\right) + |c_1|\right) + \sum_{j =
  2}^{M-2} \frac{\gamma_0 e^{-at}}{\Omega} \left(|c_j| +
|c_{j+1}|\right) + \frac{\gamma_0 e^{-at}}{\Omega}|c_{M-1}|\\ &\leq
\frac{\gamma_0 e^{-at}}{\Omega} \left(\Omega + 3 \sum_{i=1}^{M-1}
|c_i|\right)\\ &\leq \frac{\gamma_0 e^{-at}}{\Omega} \left(\Omega + 3
|\mathbf{c}|\right)\\ &= \gamma_0 e^{-at} \left(1 + \frac{3}{\Omega}
|\mathbf{c}|\right)
\end{align}

\vskip8pt

Thus, $|\mathbf{g}(t,\mathbf{c})| \leq \gamma_0 e^{-at} \left(1 +
\frac{3}{\Omega} |\mathbf{c}|\right)$, and for a given choice of $r >
0$, we may define $\eta_r(t) := \gamma_0 e^{-at} \left(1 + \frac{3
  r}{\Omega}\right)$.  From the exponential form of $\eta_r(t)$, it is
clear that $\lim_{t \to \infty} \int_t^{t+1} \eta_r(s) ds
= 0$.  Moreover, not only does there exist a single choice of $r > 0$
that produces a suitable $\eta_r(t)$, but \textit{any} choice of $r$
produces a suitable $\eta_r(t)$.

\vskip8pt

From Theorem 4.6 in~(\citealt{STRAUSS1967}), we may conclude that for
any $T_0 \geq 0$, there exists a $\delta_0 > 0$ such that if $t_0 \geq
T_0$ and $|\mathbf{c}^P(t_0)| \leq \delta_0$, then the solution of the
perturbed problem, $\mathbf{c}^P(t)$, passing through $(t_0,
\mathbf{c}^P(t_0))$ converges to zero in magnitude as $t \to \infty$.
Here the proof of convergence holds for any sufficiently smooth
function $\gamma(t) \to 0$.  Given Eq.~\ref{eq:CLONEEQ2} truncated at
an arbitrarily large threshold $M$, all $c_k$ decline with the
decaying thymic export as $t \to \infty$. While the total cell count
is preserved by proliferation driving all cells above the truncation
threshold and out of the truncated system through truncation errors,
the mean-field approximation breaks down at the limit $\gamma(t) / \mu
\to 1 / \Omega \ll 1$, and Eq.~\ref{eq:CLONEEQ2} no longer accurately
describes the real biology.  Nonetheless, our analysis here describes
the decline of the number of T-cell clones with decaying $\gamma (t)$
as $t \to \infty$, before the mean-field approximation breaks down.

\section{Computation of Expected Sample Clonal Size Distribution}
\label{sec:APP4}

\noindent In this section, we detail the derivation of
Eq.~\ref{eq:FINALEXPECTATION}, the explicit expression for
$\mathbb{E}[c_k^Y]$.  We begin with Eq.~\ref{eq:DEFEXPECTATION},

\begin{align} \label{eq:DEFEXPECTATION2}
\mathbb{E}[c_k] = \sum_{j=1}^R j P\left(c_k^Y = j\right).
\end{align}

\noindent Each term $P\left(c_k^Y = j\right)$ in
Eq.~\ref{eq:DEFEXPECTATION2} can itself be expanded as a sum over all
the ways to choose the $j$ clones that are of size $k$.  For a sample
containing exactly $Z$ clones of size $k$, we introduce the following
$Z$-tuple notation, for $Z \in \mathds{N}$:

\begin{align}
I_Z := \{\vec{i_Z} = (i_1,i_2,\ldots,i_Z): i_j \in \{1, 2, \ldots, R\},
i_j < i_{j+1} \mbox{ for all $j$}\}.
\end{align}

\noindent where $\vec{i_Z}$ lists the indices of all the sample clones
consisting of precisely $k$ cells.  Additionally, let $y_i$ denote the
size of the $i$-th ordered sample clone, so that $y_{i_1} = y_{i_2} =
\cdots = y_{i_Z} = k$, but no other sample clone consists of $k$
cells.  Note that in $\vec{i_Z}$, clones are listed in numerical
order, due to the assumption $i_j < i_{j+1}$, in order to avoid
repetition (e.g., in $I_2$, $(i_1,i_2)$ should be indistinct from
$(i_2,i_1)$, and this pair should not be counted twice, as the
significance is in which clone numbers are listed at all, and not the
order in which they are written.)  With this, let $P(\vec{i_Z},k)$
denote the probability that there are precisely $Z$ clones of size $k$
in the sample, and that their clone numbers are listed in the vector
$\vec{i_Z}$.  Additionally, for $s \in \mathbb{N}$, denote by $I_{Z,s}
\subset I_Z$ the collection of all $\vec{i_Z} \in I_Z$ such that
$i_{z^*} = s$ for some $z^* \in \{1,2,\cdots,Z\}$.  Essentially, we
are imposing the assumption that the $s$-th clone specifically belongs
somewhere in the list $\vec{i_{Z,s}}$.  Explicitly, we may write
$I_{Z,s}$ as:

\begin{align}
I_{Z,s} = \{\vec{i_{Z,s}} = (i_1,\ldots,i_{z^*-1},i_{z^*} =
s,i_{z^*+1},\ldots,i_Z): i_j \in \{1, 2, \ldots, R\}, i_j < i_{j+1}
\mbox{ for all $j$ }\}.
\end{align}

\noindent We define $P(\vec{i_{Z,s}},k)$ as the probability that there
are precisely $Z$ clones of size $k$, with clone numbers listed in
$\vec{i_{Z,s}}$, recalling that the $s$-th clone is in the list.  We
may further simplify Eq.~\ref{eq:DEFEXPECTATION2} with this notation,
rearranging sums by strategically regrouping clone size distributions
that share a common size $k$ clone.

\begin{align}
\mathbb{E}[c_k] &= \sum_{j=1}^R j P(c_k^Y = j), \label{eq:EXPECTATION1}\\
&= \sum_{j=1}^R j \left(\sum_{\vec{i_j} \in I_j} P(\vec{i_j},k)\right),\label{eq:EXPECTATION2}\\
&= \sum_{s=1}^R \left(\sum_{j=1}^R \sum_{\vec{i_{j,s}} \in I_{j,s}} P(\vec{i_{j,s}},k)\right),\label{eq:EXPECTATION3}\\
&= \sum_{s = 1}^R P(y_s = k),\label{eq:EXPECTATION4}
\end{align}

\noindent The terms of the final sum in Eq.~\ref{eq:EXPECTATION4} give
the probability that the $s$-th clone is of size $k$, independent of
any other information about the sampling.  This probability is easy to
compute, and given by:

\begin{align} \label{eq:EASYPROBABILITY}
P(y_s = k) = \frac{1}{\binom{N}{Y}} \binom{n_s}{k} \binom{N-n_s}{Y-k}.
\end{align}

\noindent Inserting Eq.~\ref{eq:EASYPROBABILITY} into
Eq.~\ref{eq:EXPECTATION4}, we obtain a simple expression for the
expected sample clone size distribution:

\begin{align} \label{eq:SAMPLECLONEDIST}
\mathbb{E}[c_k] &= \sum_{s=1}^R \frac{1}{\binom{N}{Y}} \binom{n_s}{k} \binom{N-n_s}{Y-k}.
\end{align}

\noindent We can further simplify Eq.~\ref{eq:SAMPLECLONEDIST} by
recognizing that the term $\binom{n_s}{k}$ is nonzero only if $n_s
\geq k$.  We can thus rewrite Eq.~\ref{eq:SAMPLECLONEDIST} in terms of
the true clone size distribution $\{c_l^N\}_{l=1}^R$ as:

\begin{align}
\mathbb{E}[c_k] &= \sum_{l=k}^R \frac{1}{\binom{N}{Y}} c_l^N
\binom{l}{k} \binom{N-l}{Y-k}.
\end{align}


\bibliographystyle{spbasic}
\bibliography{naivetcellsbib}

\begin{thebibliography}{61}
\providecommand{\natexlab}[1]{#1}
\providecommand{\url}[1]{{#1}}
\providecommand{\urlprefix}{URL }
\expandafter\ifx\csname urlstyle\endcsname\relax
  \providecommand{\doi}[1]{DOI~\discretionary{}{}{}#1}\else
  \providecommand{\doi}{DOI~\discretionary{}{}{}\begingroup
  \urlstyle{rm}\Url}\fi
\providecommand{\eprint}[2][]{\url{#2}}

\bibitem[{Appay and Sauce(2014)}]{APPAY2014}
Appay V, Sauce D (2014) Naive {T} cells: The crux of cellular immune aging?
  Experimental Gerontology 54:90--93

\bibitem[{Bains et~al.(2009{\natexlab{a}})Bains, Antia, Callard, and
  Yates}]{BAINS2009}
Bains I, Antia R, Callard R, Yates AJ (2009{\natexlab{a}}) Quantifying the
  development of the peripheral naive {CD}4+ {T}-cell pool in humans.
  Immunobiology 113(22):5480--5487

\bibitem[{Bains et~al.(2009{\natexlab{b}})Bains, Thi{\'e}baut, Yates, and
  Callard}]{BAINS20092}
Bains I, Thi{\'e}baut R, Yates AJ, Callard R (2009{\natexlab{b}}) Quantifying
  thymic export: {C}ombining models of naive {T} cell proliferation and {TCR}
  excision circle dynamics gives an explicit measure of thymic output. The
  Journal of Immunology 183(7):4329--4336

\bibitem[{Berzins et~al.(1998)Berzins, Boyd, and Miller}]{BERZINS1998}
Berzins SP, Boyd R, Miller JF (1998) The role of the thymus and recent thymic
  migrants in the maintenance of the adult peripheral lymphocyte pool. The
  Journal of Experimental Medicine 187(11):1839--1848

\bibitem[{Bradley et~al.(2005)Bradley, Haynes, and Swain}]{BRADLEY2005}
Bradley LM, Haynes L, Swain SL (2005) {IL}-7: maintaining {T}-cell memory and
  achieving homeostasis. Trends in Immunology 26(3):172--176

\bibitem[{Brass et~al.(2014)Brass, McKay, and Scott}]{BRASS2014}
Brass D, McKay P, Scott F (2014) Investigating an incidental finding of
  lymphopenia. British Medical Journal 348:1--3

\bibitem[{Britanova et~al.(2014)Britanova, Putintseva, Shugay, Merzlyak,
  Turchaninova, Staroverov, Bolotin, Lukyanov, Bogdanova, Mamedov, Lebedev, and
  Chudakov}]{BRITANOVA2014}
Britanova OV, Putintseva EV, Shugay M, Merzlyak EM, Turchaninova MA, Staroverov
  DB, Bolotin DA, Lukyanov S, Bogdanova EA, Mamedov IZ, Lebedev YB, Chudakov DM
  (2014) Age-related decrease in {TCR} repertoire diversity measured with deep
  and normalized sequence profiling. The Journal of Immunology
  192(6):2689--2698

\bibitem[{Chao(1984)}]{CHAO1984}
Chao A (1984) Nonparametric estimation of the number of classes in a
  population. Scandanavian Journal of Statistics 11(4):265--270

\bibitem[{Chao and Lee(1992)}]{CHAO1992}
Chao A, Lee SM (1992) Estimating the number of classes via sample coverage.
  Journal of the American Statistical Association 87(417):210--217

\bibitem[{Colwell and Coddington(1994)}]{COLWELL1994}
Colwell RK, Coddington JA (1994) Estimating terrestrial biodiversity through
  extrapolation. Philosophical Transactions of the Royal Society B
  345(1311):101--118

\bibitem[{{de Boer} and Perelson(2013)}]{DEBOER2013}
{de Boer} RJ, Perelson AS (2013) Quantifying {T} lymphocyte turnover. Journal
  of Theoretical Biology 327:45--87

\bibitem[{Desponds et~al.(2015)Desponds, Mora, and Walczak}]{DESPONDS2015}
Desponds J, Mora T, Walczak A (2015) Fluctuating fitness shapes the clone-size
  distribution of immune repertoires. Proceedings of the National Academy of
  Sciences 113(2):274--279

\bibitem[{Desponds et~al.(2017)Desponds, Mayer, Mora, and
  Walczak}]{DESPONDS2017}
Desponds J, Mayer A, Mora T, Walczak AM (2017) Population dynamics of immune
  repertoires. ArXiv e-prints \eprint{1703.00226}

\bibitem[{Ewens(1972)}]{EWENS1972}
Ewens W (1972) The sampling theory of selectively neutral alleles. Theoretical
  Population Biology 3(1):87--112

\bibitem[{Fagnoni et~al.(2000)Fagnoni, Vescovini, Passeri, Bologna, Pedrazzoni,
  Lavagetto, Casti, Franceschi, Passeri, and Sansoni}]{FAGNONI2000}
Fagnoni FF, Vescovini R, Passeri G, Bologna G, Pedrazzoni M, Lavagetto G, Casti
  A, Franceschi C, Passeri M, Sansoni P (2000) Shortage of circulating naive
  {CD}8+ {T} cells provides new insights on immunodeficiency in aging. Blood
  95(9):2860--2868

\bibitem[{Fleming and Elliot(2008)}]{FLEMING2005}
Fleming DM, Elliot AJ (2008) The impact of influenza on health and health care
  utilisation of elderly people. Vaccine 32(1):S1--S9

\bibitem[{Fry and Mackall(2005)}]{FRY2005}
Fry TJ, Mackall CL (2005) The many faces of {IL}-7: from lymphopoesis to
  peripheral {T} cell maintenance. The Journal of Immunology 174(11):6571--6576

\bibitem[{Gergely(1999)}]{GERGELY1999}
Gergely P (1999) Drug-{I}nduced {L}ymphopenia. Drug Safety 21(2):91--100

\bibitem[{Ginaldi et~al.(2001)Ginaldi, Loreto, Corsi, Modesti, and {de
  Martinis}}]{GINALDI2001}
Ginaldi L, Loreto MF, Corsi MP, Modesti M, {de Martinis} M (2001)
  Immunosenescence and infectious diseases. Microbes and Infection
  3(10):851--857

\bibitem[{Globerson and Effros(2000)}]{GLOBERSON2000}
Globerson A, Effros RB (2000) Aging of lymphocytes and lymphocytes in the aged.
  Immunology Today 21(10):515--521

\bibitem[{Goronzy et~al.(2007)Goronzy, Lee, and Weyland}]{GORONZY2007}
Goronzy JJ, Lee WW, Weyland CM (2007) Aging and {T}-cell diversity.
  Experimental Gerontology 42(5):400--406

\bibitem[{Goronzy et~al.(2015{\natexlab{a}})Goronzy, Fang, Cavanagh, Qi, and
  Weyand}]{GORONZY2015b}
Goronzy JJ, Fang F, Cavanagh MM, Qi Q, Weyand CM (2015{\natexlab{a}}) Na\"ive
  {T} cell maintenance and function in human aging. Journal of Immunology
  194(9):4073--4080

\bibitem[{Goronzy et~al.(2015{\natexlab{b}})Goronzy, Qi, Olshen, and
  Weyland}]{GORONZY2015}
Goronzy JJ, Qi Q, Olshen RA, Weyland CM (2015{\natexlab{b}}) High-throughput
  sequencing insights into {T}-cell receptor diversity in aging. Genome
  Medicine 7:1--3

\bibitem[{Goyal et~al.(2015)Goyal, Kim, Chen, and Chou}]{GOYAL2015}
Goyal S, Kim S, Chen ISY, Chou T (2015) Mechanisms of blood homeostasis:
  lineage tracking and a neutral model of cell populations in rhesus macaques.
  BMC Biology 13(85):1--14

\bibitem[{Grossman et~al.(2015)Grossman, Ellsworth, Campian, Wild, Herman,
  Laheru, Brock, Balmanoukian, and Ye}]{GROSSMAN2015}
Grossman SA, Ellsworth S, Campian J, Wild AT, Herman JM, Laheru D, Brock M,
  Balmanoukian A, Ye X (2015) Survival in patients with severe lymphopenia
  following treatment with radiation and chemotherapy for newly diagnosed solid
  tumors. Journal of the National Comprehensive Cancer Network
  13(10):1225--1231

\bibitem[{Gruver et~al.(2007)Gruver, Hudson, and Sempowski}]{GRUVER2007}
Gruver A, Hudson L, Sempowski J (2007) Immunosenescence of aging. The Journal
  of Pathology 211(2):144--156

\bibitem[{Hapuarachchi et~al.(2013)Hapuarachchi, Lewis, and
  Callard}]{HAPUARACHCHI2013}
Hapuarachchi T, Lewis J, Callard RE (2013) A mechanistic mathematical model for
  naive {CD}4 {T} cell homeostasis in healthy adults and children. Frontiers in
  Immunology 4(366):1--6

\bibitem[{Jenkins et~al.(2009)Jenkins, Chu, McLachlan, and Moon}]{JENKINS2009}
Jenkins MK, Chu HH, McLachlan JB, Moon JJ (2009) On the composition of the
  preimmune repertoire of {T} cells specific for peptide-major
  histocompatibility ligands. Annual Review of Immunology 28:275--294

\bibitem[{Johnson et~al.(2012)Johnson, Yates, Goronzy, and Antia}]{JOHNSON2012}
Johnson PLF, Yates AJ, Goronzy JJ, Antia R (2012) Peripheral selection rather
  than thymic involution explains sudden contraction in naive {CD4} {T-cell}
  diversity with age. PNAS 109(52):21432--21437

\bibitem[{Johnson et~al.(2014)Johnson, Goronzy, and Atia}]{JOHNSON2014}
Johnson PLF, Goronzy JJ, Atia R (2014) A population biological approach to
  understanding the maintenance and loss of the {T-cell} repertoire during
  aging. Immunology 142(2):167--175

\bibitem[{Laydon et~al.(2015)Laydon, Bangham, and Asquith}]{LAYDON2015}
Laydon DJ, Bangham CRM, Asquith B (2015) Estimating {T}-cell repertoire
  diversity: limitations of classical estimators and a new approach.
  Philosophical Transactions of the Royal Society B 370(1675):1--11

\bibitem[{Lythe et~al.(2016)Lythe, Callard, Hoare, and
  Molina-Par{'\i}s}]{LYTHE2016}
Lythe G, Callard RE, Hoare RL, Molina-Par{'\i}s C (2016) How many {TCR}
  clonotypes does a body maintain? Journal of Theoretical Biology 389:214--224

\bibitem[{Mason(1998)}]{MASON1998}
Mason D (1998) A very high level of crossreactivity is an essential feature of
  the {T}-cell receptor. Trends in Immunology 19(9):395--404

\bibitem[{McElhaney and Dutz(2008)}]{MCELHANEY2008}
McElhaney JA, Dutz JP (2008) Better influenza vaccines for older people: What
  will it take? The Journal of Infectious Diseases 198(5):632--634

\bibitem[{Mehr et~al.(1996)Mehr, Perelson, Fridkis-Hareli, and
  Globerson}]{MEHR1996}
Mehr R, Perelson AS, Fridkis-Hareli M, Globerson A (1996) Feedback regulation
  of {T} cell development: manifestations in aging. Mechanisms of Ageing and
  Development 91(3):195--210

\bibitem[{Mehr et~al.(1997)Mehr, Perelson, Fridkis-Harelic, and
  Globersond}]{MEHR1997}
Mehr R, Perelson AS, Fridkis-Harelic M, Globersond A (1997) Regulatory feedback
  pathways in the thymus. Immunology Today 18(12):581--585

\bibitem[{Metcalf(1963)}]{METCALF1963}
Metcalf D (1963) The autonomous behavior of normal thymus grafts. Australian
  Journal of Experimental Biology and Medical Sciences 41:437--444

\bibitem[{Mora and Walczak(2016)}]{MORA2016}
Mora T, Walczak A (2016) Quantifying lymphocyte receptor diversity. ArXiv
  e-prints \eprint{1604.00487}

\bibitem[{Moro-Garc{\'\i}a et~al.(2013)Moro-Garc{\'\i}a, Arias, and
  L{\'o}pez-Arrea}]{MORO-GARCIA2013}
Moro-Garc{\'\i}a MA, Arias RA, L{\'o}pez-Arrea C (2013) When aging reaches
  {CD}4+ {T}-cells: {P}henotypic and functional changes. Frontiers in
  Immunology 4(107):1--12

\bibitem[{Morris et~al.(2014)Morris, Caruso, Buscot, Fischer, Hancock, Maier,
  Meiners, M{\"u}ller, Obermaier, Prati, Socher, Sonnemann, W{\"a}schke, Wubet,
  Wurst, and Rillig}]{MORRIS2014}
Morris EK, Caruso T, Buscot F, Fischer M, Hancock C, Maier TS, Meiners T,
  M{\"u}ller C, Obermaier E, Prati D, Socher SA, Sonnemann I, W{\"a}schke N,
  Wubet T, Wurst S, Rillig MC (2014) Choosing and using diversity indices:
  {I}nsights for biological applications from the {G}erman {B}iodiversity
  {E}xploratories. Ecology and Evolution 4:3514--3524

\bibitem[{Murphy(2012)}]{JANEWAY2012}
Murphy K (2012) Immunobiology. Garland Science, Taylor and Francis Group, LLC

\bibitem[{Murray et~al.(2003)Murray, Kaufmann, Hodgkin, Lewin, Kelleher,
  Davenport, and Zaunders}]{MURRAY2003}
Murray JM, Kaufmann GR, Hodgkin PD, Lewin SR, Kelleher AD, Davenport MP,
  Zaunders JJ (2003) Naive {T}-cells are maintained by thymic output in early
  ages but by proliferation without phenotypic change after age 20. Immunology
  and Cell Biology 81(6):487--495

\bibitem[{Naylor et~al.(2005)Naylor, Li, Vallejo, Lee, Koetz, Bryl, Witkowski,
  Fulbright, Weyand, and Goronzy}]{NAYLOR2005}
Naylor K, Li G, Vallejo AN, Lee WW, Koetz K, Bryl E, Witkowski J, Fulbright J,
  Weyand CM, Goronzy JJ (2005) The influence of age on {T} cell generation and
  {TCR} diversity. The Journal of Immunology 174(11):7446--7452

\bibitem[{Poland et~al.(2010)Poland, Langley, Michel, {Van Damme}, and
  Wicker}]{POLAND2010}
Poland GA, Langley J, Michel J, {Van Damme} P, Wicker S (2010) A global
  prescription for adult immunization: Time is catching up with us. Vaccine
  28(44):7137--7139

\bibitem[{Qi et~al.(2014)Qi, Liu, Cheng, Glanville, Zhang, Lee, Olshen, Weyand,
  Boyd, and Goronzy}]{QI2014}
Qi Q, Liu Y, Cheng Y, Glanville J, Zhang D, Lee JY, Olshen RA, Weyand CM, Boyd
  SD, Goronzy JJ (2014) Diversity and clonal selection in the human {T}-cell
  repertoire. Proceedings of the National Academy of Sciences
  111(36):13139--13144

\bibitem[{Reynolds et~al.(2013)Reynolds, Coles, Lythe, and
  Molina-Par{\'\i}s}]{REYNOLDS2013}
Reynolds J, Coles M, Lythe G, Molina-Par{\'\i}s C (2013) Mathematical model of
  naive {T} cell division and {IL}-7 survival thresholds. Frontiers in
  Immunology 4(434):1--13

\bibitem[{Ribeiro and Perelson(2007)}]{RIBEIRO2007}
Ribeiro RM, Perelson AS (2007) Determining thymic output quantitatively:
  {U}sing models to interpret experimental {T}-cell receptor excision circle
  ({TREC}) data. Immunological Reviews 216(1):21--34

\bibitem[{Salam et~al.(2013)Salam, Rane, Das, Faulkner, Gund, Kandpal, Lewis,
  Prabhu, Ranganathan, Durdik, George, Rath, and Bal}]{SALAM2013}
Salam N, Rane S, Das R, Faulkner M, Gund R, Kandpal U, Lewis V, Prabhu HMS,
  Ranganathan V, Durdik J, George A, Rath S, Bal V (2013) T cell ageing:
  Effects of age on development, survival, \& function. Indian Journal of
  Medical Research 138(5):595--608

\bibitem[{Steinmann et~al.(1985)Steinmann, Klaus, and
  M{\"u}ller-Hermelink}]{STEINMANN1985}
Steinmann G, Klaus B, M{\"u}ller-Hermelink H (1985) The involution of the
  ageing human thymic epithelium is independent of puberty. Scandanavian
  Journal of Immunology 22(5):563--575

\bibitem[{Steinmann(1986)}]{STEINMANN1986}
Steinmann GG (1986) The {H}uman {T}hymus, Current Topics in Pathology, vol~75,
  Springer Berlin Heidelberg, chap Changes in {T}he {H}uman {T}hymus {D}uring
  {A}ging, pp 43--88

\bibitem[{Strauss and Yorke(1967)}]{STRAUSS1967}
Strauss A, Yorke JA (1967) Perturbation theorems for ordinary differential
  equations. Journal of Differential Equations 3(1):15--30

\bibitem[{Tan et~al.(2001)Tan, Dudl, Le{R}oy, Murray, Sprent, Weinberg, and
  Surh}]{TAN2001}
Tan JT, Dudl E, Le{R}oy E, Murray R, Sprent J, Weinberg KI, Surh CD (2001)
  {IL}-7 is critical for homeostatic proliferation and survival of naive {T}
  cells. Proceedings of the National Academy of Sciences 98(15):8732--8737

\bibitem[{Thomas-Crussels et~al.(2012)Thomas-Crussels, McElhaney, and
  Aguado}]{THOMAS2012}
Thomas-Crussels J, McElhaney JE, Aguado MT (2012) Report of the ad-hoc
  consultation on aging and immunization for a future who research agenda on
  life-course immunization. Vaccine 40(32):6007--6012

\bibitem[{Trepel(1974)}]{TREPEL1974}
Trepel F (1974) Number and distribution of lymphocytes in man. Klinische
  Wochenschrift 52(11):511--515

\bibitem[{Vivien et~al.(2001)Vivien, Benoist, and Mathis}]{VIVIEN2001}
Vivien L, Benoist C, Mathis D (2001) T lymphocytes need {IL}-7 but not {IL}-4
  or {IL}-6 to survive in vivo. International Immunology 13(6):763--768

\bibitem[{Vrisekoop et~al.(2008)Vrisekoop, den Braber, de~Boer, Ruiter,
  Ackermans, van~der Crabben, Schrijver, Spierenburg, Sauerwein, Hazenberg,
  de~Boer, Miedema, Borghans, and Tesselaar}]{VRISEKOOP2008}
Vrisekoop N, den Braber I, de~Boer AB, Ruiter AFC, Ackermans MT, van~der
  Crabben SN, Schrijver EHR, Spierenburg G, Sauerwein HP, Hazenberg MD, de~Boer
  RJ, Miedema F, Borghans JAM, Tesselaar K (2008) Sparse production but
  preferential incorporation of recently produced naive {T}-cells in the human
  peripheral pool. Proceedings of the National Academy of Sciences
  105(16):6115--6120

\bibitem[{Westera et~al.(2015)Westera, van Hoeven, Drylewicz, Spierenburg, van
  Velzen, de~Boer, Tesselaar, and Borghans}]{WESTERA2015}
Westera L, van Hoeven V, Drylewicz J, Spierenburg G, van Velzen JF, de~Boer RJ,
  Tesselaar K, Borghans JAM (2015) Lymphocyte maintenance during healthy aging
  requires no substantial alterations in cellular turnover. Aging Cell
  14(2):219--227

\bibitem[{Westermann and Pabst(1990)}]{WESTERMANN1990}
Westermann J, Pabst R (1990) Lymphocyte subsets in the blood: {A} diagnostic
  window on the lymphoid system? Immunology Today 11(11):406--410

\bibitem[{Wick et~al.(2000)Wick, D{\"u}rr, Berger, Blasko, and
  Grubeck-Loebenstein}]{WICK2000}
Wick G, D{\"u}rr PJ, Berger P, Blasko I, Grubeck-Loebenstein B (2000) Diseases
  of aging. Vaccine 18(16):1567--1583

\bibitem[{Yagger et~al.(2008)Yagger, Ahmed, Lanzer, Randall, Woodland, and
  Blackman}]{YAGGER2008}
Yagger EJ, Ahmed M, Lanzer K, Randall TD, Woodland DL, Blackman MA (2008)
  Age-associated decline in {T} cell repertoire diversity leads to holes in the
  repertoire and impaired immunity to the influenza virus. The Journal of
  Experimental Medicine 205(3):711--723

\bibitem[{Yates(2014)}]{YATES2014}
Yates AJ (2014) Theories and quantification of thymic selection. Frontiers in
  Immunology 5(13):1--15

\end{thebibliography}

\end{document}